\documentclass[12pt]{article}
\usepackage{amsmath,epsfig}
\newcommand{\beq}{\begin{eqnarray}}
\newcommand{\eeq}{\end{eqnarray}}
\begin{document}
\title{2004 Review of Light Cone Field Theory}
\author{Leonard S. Kisslinger$^\dagger$\\
    Department of Physics, Carnegie Mellon University, Pittsburgh, PA 15213}
\maketitle
\indent
 

\vspace{0.5 in}

\noindent
PACS Indices: 11.30.Cp,12.38.Lg,12.39.Ki,13.40.Gp,13.25.Hw \\
\vspace{2 mm}
$^\dagger$ email: kissling@andrew.cmu.edu

\vspace{2cm}

{\bf CONTENTS}

\vspace{5mm}

   {\bf 1 Introduction}
\vspace{5mm}

   {\bf 2 Light Cone Quantum Mechanics}
\vspace{5mm}

   {\bf 3 Light Cone Field Theory}
\vspace{5mm}

   {\bf 4  B-S Equation in a Light Cone Representation and Pion Form 
Factor}
\vspace{5mm}

   {\bf 5 Light Cone Representation of the Quark Schwinger-Dyson Equation}
\vspace{5mm}

   {\bf 6 Deeply Virtual Compton Scattering and Skewed Parton Distributions}
\vspace{5mm}

   {\bf 7 Rare $B\to K\ell^+\ell^-$ Decays}
\vspace{5mm}

   {\bf 8 Meson light Cone Wave Functions}
\vspace{5mm}

   {\bf 9 Factorization and Experimental Study of Light Cone Wave Functions}

\section{Introduction}

\hspace{.5cm}

   Light cone quantization was introduced by Dirac in his exploration\cite{dir}
of possible representations of the Poincar$\acute{e}$ Group for 
relativistic formulations of quantum mechanics. The
most obvious application in nuclear and particle physics is to high momentum
transfer processes for hadrons and nuclei, since in a light cone representation
a diagonal Lorenz boost operator can be defined, while in the standard 
instant form representation Lorentz boosts of bound systems are difficult to 
formulate.

   There have been several reviews of light cone quantization for Hamiltonian
Dynamics with various applications to bound states (see Ref.\cite{bpp} for 
references) and the light cone representations of bound states in 
supersymmetric theories\cite{hpt02}. The light cone community holds an annual
meeting, and this review is based in part on developments in this area
presented and discussed at the International Workshop Light cone 
2002\cite{lc02}

   The light cone formulation of Quantum Chromodynamics (QCD) field  theory, 
the presently accepted theory of strong interactions, is particularly  
important, since only by treating processes from low momentum transfer 
to high momentum transfer can one explore the transition from nonperturbative 
to perturbative regions, which is still an unsolved problem. Light cone 
formulations are important for such studies.  E.g., in one of the early 
applications of light cone formulations to form factors it was 
shown\cite{jk86} that with a model that predicts that nonperturbative 
processes dominate the pion form factor until momentum transfer 
$Q^2 \simeq 4 GeV^2$, and that at
$Q^2 >$ 5 GeV$^2$ perturbative QCD theory might begin to be applicable, 
it was found in an instant form treatment of the same 
model\cite{ils84} that only at much higher values of $Q^2$ can perturbative 
QCD be used. Light cone treatments of the pion form factor are reviewed below.

In the present review we shall 
center on applications to processes involving either high momentum transfer 
or both high and low momentum transfer.

\section{Light Cone Quantum Mechanics}

\hspace{.5cm}
   
   In this section we review the light cone formulation of relativistic
quantum mechanics. Quantum mechanics is based on states and operators, 
which must tranform correctly under Lorentz transformations in a relativistic
theory. The Poincar$\acute{e}$ group of translations, rotations, and
Lorentz transformations in the standard instant form has serious problems
for Lorentz transformations, and as we shall see this problem is much
less serious in the light cone representation.

   These problems are more serious in field theories, which we review in the 
following section, and a number of the concepts which arise in quantum
mechanics apply with modifications in field theories.

\subsection{Lorentz Transformations and the Poincar$\acute{e}$ Group}

   The invariant distance, $ds^2$ is defined in terms of the four position,
$x^\mu$ by the metric tensor, $g^{\mu\nu}$, by
\beq
\label{metric}
   ds^2 & = & g_{\mu\nu}x^\mu x^\nu,
\eeq
where the Greek indices run (1,...4). An inhomogeneous Lorentz transformation
of the position 4-vector is given by
\beq
\label{lorentz}
     x^{'\mu} & = & a^\mu + \Lambda^\mu_\nu x^\nu \; ,
\eeq
where $a_\mu$ is a constant and $\Lambda^\mu_\nu$ is the constant Lorentz
matrix, which satisfies
\beq
\label{LL}
    \Lambda^\mu_\nu  \Lambda^{\nu\lambda} & = & g^{\mu\lambda}.
\eeq

  The ten generators of infinitesimal Lorentz transformations are:
\beq
\label{Pgens}
      P^\mu && {\rm are \: four \: momentum\: operators} \\ \nonumber
      M^{\mu\nu} && {\rm give \: six\: independent\: angular\: momentum\: 
operators}
\eeq
with $M^{\mu\nu} = -M^{\nu\mu}$.

These ten operators form the Poincar$\acute{e}$ group, which satisfy the
the commutation rules (Poincar$\acute{e}$ Algebra)
\beq
\label{Palgebra}
    [P^\mu,P^\nu] &=& 0 \\ \nonumber
 [M^{\mu\nu},P^{\lambda}] &=& i(g^{\nu\lambda}P^\mu -g^{\mu\lambda}P^\nu) \\
\nonumber
   [M^{\mu\nu},M^{\lambda\sigma}] &=& i(g^{\mu\sigma} M^{\nu\lambda}+
g^{\nu\lambda} M^{\mu\sigma} -g^{\mu\lambda} M^{\nu\sigma}-
g^{\nu\sigma} M^{\mu\lambda})
\eeq

In the next two subsections we consider the instant form and light cone 
representations.

\subsubsection{Instant Form for Dirac Particles}

  The instant form is the one commonly used in relativistic quantum mechanics:
\beq
\label{instant}
    x^\mu &= & (x^o,\vec{x}) \: = \: (t,x^1,x^2,x^3 = z) \\ \nonumber
    g^{\mu\mu} &=& (1,-1,-1,-1) \\
    g^{\mu\nu} &=& 0 \; {\rm for} \; \mu \neq \nu,
\eeq
with $\mu$ = (0,1,2,3).

Let us first consider a free Dirac Particle.  The ten generators 
(c= h/$2\pi$ =1) at t=0 are
\beq
\label{instgen}
    P^\mu & = & (P^0,\vec{P})\\ \nonumber
   J_i &=& \epsilon_{ijk}M^{jk}  \; = \; (\vec{r} \times \vec{P})_i + S_i \\
\nonumber
   K_i &=& M_{0i} \; = \; \sigma_{0i},
\eeq
where $\sigma_{\mu\nu} = i[\gamma_\mu,\gamma_\nu]/2 $. $P^\mu$ and $\vec{S}$
are the usual momentum and internal spin operators. Wave functions transform
under the group by
\beq
\label{wf}
  {\rm translations} \; \; \psi'(x + X) &=& e^{i\vec{P} \cdot \vec{X}}\psi(x)\\
    {\rm rotations} \; \; \psi'(\vec{r} \rightarrow \vec{r'}(\theta \hat{n}))
    &=& e^{i\theta \vec{J}\cdot\hat{n}}\psi(x)\\
{\rm Lorentz \; transformations} \; \; \psi'(x^{\mu'}) &=& e^{i\omega \vec{K}
 \cdot \hat{n}}\psi(x^\mu),
\eeq
with a rotation of $\theta$ about the $\hat{n}$ axis and a Lorentz boost of
tanh$\omega$ = v in the $\hat{n}$ direction.

For a system of interacting fermions in the instant form the boost operator
$\vec{K}$ contains the interaction potential. There has been a great deal of
literature on this subject, starting with Foldy's expansion in c$^{-2}$
\cite{foldy}. We do not discuss this further here.

\subsubsection{Light Cone Form for Dirac Particles} 

   The light cone form is defined by
\beq
\label{lc}
     x^\mu &= & (x^+,\vec{x}_\perp,x^-) \: = \: (x^+,x^1,x^2,x^-) \\ \nonumber
     x^{\pm} &=& (t \pm z)/\sqrt{2} \\ \nonumber
    g^{\mu\nu} &=& \left(\begin{array}{clcr}
        0 &  0 &  0 &  1 \\
        0 & -1 &  0 &  0 \\
        0 &  0 & -1 &  0 \\
        1 &  0 &  0 &  0     \end{array} \right)
\eeq
The Poincar$\acute{e}$ generators are
\beq
\label{lcP}
      P^\mu &=& (P^+,\vec{P}_\perp,P^-) \\ \nonumber
      P^{\pm} &=& (P^0 \pm P^3) \\ \nonumber
      \vec{P}_\perp &=& (P_1,P_2)
\eeq
momentum operators and
\beq
\label{lcM}
   M_{\mu\nu} &=& \left(\begin{array}{clcr}
       0   &  -F_1 & -F_2 &  K_3 \\
       F_1 &   0   &  J_3 &  E_1 \\
       F_2 &  -J_3 &   0  &  E_2 \\
      -K_3 &  -E_1 & -E_2 &   0      \end{array} \right)
\eeq
angular momentum operators, with
\beq
\label{FK}
    F_1 &=& (K_1 -J_2)/\sqrt{2} \; \; \; F_2 = (K_2 +J_1)/\sqrt{2} \\ \nonumber
    E_1 &=& (K_1 +J_2)/\sqrt{2} \; \; \; E_2 =  (K_2 -J_1)/\sqrt{2} \: .
\eeq

   In this light cone form one can choose seven generators to be independent
of the interaction (often called ``good'' generators), such as (see 
\cite{dir,le78,ba79})
\beq
\label{good}
  Good:\: \: P^+, \: P^1, \: P^2, \: E_1, \: E_2, \: K_3, \: J_3, &&
\eeq
so that one can carry out a Lorentz boost in one direction without involving
interactions. This is also true for field theory, as we shall see, and 
is most significant, since interactions generally produce particles. The
other three generators (the ``bad'' generators)
\beq
   Bad:\: \: F_1, \: F_2, \: P^- &&
\eeq
contain the interactions. Therefore, although a boost in the z direction is
interaction free, rotations about the x and y axes are not; therefore
prescriptions for total angular momentum for composite states are needed.
See, e.g., Ref\cite{kt80}.

\subsection{Light Front Hamiltonian Dynamics}

   One method for constructing a relativistic quantum theory for interacting
particles with an interaction-independent Lorentz boost follows the
Bakamjian-Thomas idea\cite{bt53} of using M, the invariant mass 
operator. See Ref.\cite{kp} for a review and references.

   The energy operator in this method is $P^-$
\beq
\label{lfhd}
   P^- & = & \frac{M^2 + P_\perp^2}{P^+}.
\eeq
The method of obtaining solutions to the Hamiltonian eigenvalue problem,
with the Hamiltonian operator $H$ containing the interactions,
\beq
\label{hev}
    H|\Psi> &=& \frac{P^-}{2}|\Psi>
\eeq
for the states and wave functions by a Fock state expansion is discussed
in detail in Ref.\cite{bpp}, with discussion of applications to QCD and
QED. There is an extensive discussion of applications for relativistic 
corrections in nuclear physics in Ref.\cite{kp}. In the present review
we concentrate on applications of light cone field theory, which is
discussed next.

\section{Light Cone Field Theory}
   In this section we review the light cone formalism for quantum field 
theory. One of our main objectives is to discuss the derivation of the
ten Poincar$\acute{e}$ generators, with an interaction-free Lorentz
boost operator. First we review the instant form procedure for deriving
the Poincar$\acute{e}$ generators.

\subsection{Poincar$\acute{e}$ Generators in Instant Form Field Theory}

   The starting point in a quantum field theory is the Lagrangian density,
${\cal L}(x)={\cal L}(\psi^\alpha,\partial^\mu \psi^\alpha)$,
with the fields $\psi^\alpha$, the conjugate momenta $ \pi_\beta(\vec{x},t)
= \partial {\cal L}(x)/(\partial_0  \psi_\beta(\vec{x},t))$, and the
commutation rules of the fields,
\beq
\label{commutation}
   [\psi^\alpha(\vec{x},t),\pi _\beta(\vec{x}',t)] &=& i\delta_{\alpha,\beta}
 \delta(\vec{x}-\vec{x}').
\eeq
  
   The Poincar$\acute{e}$ generators are obtained from the energy momentum
tensor, which can be derived from
\beq
\label{enmom}
     T^{\mu\nu} &=& \frac{\partial {\cal L}(x)}{\partial (\partial_\mu
 \psi^\alpha)}\partial_\nu \psi^\alpha -g^{\mu\nu} {\cal L}(x).
\eeq
E.g., the Lagrangian density for pure glue QCD is
\beq
\label{glue}
  {\cal L}^{(glue)} & = & \frac{1}{4} G \cdot G
\eeq
with
\beq
\label{G}
    G_{\mu\nu} & = & \partial_\mu A_\nu -  \partial_\nu A_\mu
-i g [A_\mu,A_\nu]\\
    A_\mu & = & A_\mu^n \lambda^n/2 \nonumber
\eeq
where $\lambda^n$ are the eight SU(3) Gell-Mann matrices, 
$([\lambda_a,\lambda_b]=2if_{abc}\lambda_c)$. The corresponding energy
momentum tensor is given by
\beq
\label{emt}
T^{(glue) \, \mu \nu} &=& \sum_a(G^{\mu \alpha}_a G_{\alpha a}^\nu
-\frac{1}{4}g^{\mu \nu} G^{\alpha \beta}_a G_{\alpha \beta a}).
\eeq

The momentum operators are 
\beq
\label{momen}
         P^\mu(x) &=& T^{0\mu}(x)
\eeq
and the Hamiltonian density is 
\beq
\label{ham}
       {\cal H}(x) &=& T^{00}.
\eeq
The form of the $M^{\mu\nu}$ tensor depends on the specific theory.
We consider two examples next.

\subsubsection{Poincar$\acute{e}$ Operators for Scalar Fields}

  For a scalar field with the Lagrangian density
\beq
\label{scalar}
     {\cal L} &=& \frac{1}{2}(\partial_\mu \phi(x))^2 -\frac{1}{2}\phi(x)^2
 +{\cal L}_{int}(x),
\eeq
one finds
\beq
\label{scapgroup}
    {\cal H} &=& T^{00} \; = \; (\partial_0 \phi(x))^2 -{\cal L}_{int}(x)\\
\nonumber
     M^{\mu\nu} &=& \int d^3 x[x^\nu T^{0\mu}-x^\mu T^{0\nu}] \\ \nonumber
     J^i &=& \epsilon^{ijk} M_{jk} \\ \nonumber
     K^i &=& M^{i0} \; = \; \int d^3 x^i {\cal H},
\eeq
where the boost operator $K^i$ has been evaluated at t=0. It is clear that
the $K^i$ contain the interactions.

\subsubsection{Poincar$\acute{e}$ Operators for Dirac Fields}

The Larangian for interacting Dirac particles is
\beq
\label{diracL}
   {\cal L}(x) &=& \frac{i}{2}[\bar{\psi}(x)\gamma^\mu\partial_\mu \psi(x)
 -(\partial_\mu\bar{\psi}(x))\gamma^\mu\psi(x)]  -m \bar{\psi}(x) \psi(x)
 + {\cal L}_{int}(x).
\eeq
One finds that the internal spin leads to an additional term in the angular
momentum operators:
\beq
\label{dmom}
  M^{\mu\nu} &=& \int d^3 x [x^\nu T^{0\mu}-x^\mu T^{0\nu} 
-\frac{i}{2} \frac{\partial {\cal L}}{\partial(\partial_0 \psi)}
\sigma^{\mu\nu}\psi].
\eeq
Using $K^i = M^{0i}$, at t=0  one obtains for the boost operator
\beq
\label{dboost}
  \vec{K} &=& \int d^3 x[\frac{1}{2}\vec{x}{\cal H}(x)-\frac{i}{2}
\bar{\psi}(x)\vec{\gamma}\psi(x)] \; .
\eeq
As expected, the boost operator contains the interactions.  Thus the state
obtained by a boost with velocity $\bar{v}= v\hat{n}$ of the state 
$|\psi_o>$,
\beq
\label{bstate}
 |\psi_{\bar{v}}> &=& e^{i \omega \vec{K}\cdot\hat{n}} |\psi_0>,
\eeq
will be quite different form the state at rest, since the boost $\vec{K}$
contains fields which create and destroy particles. This is a serious
problem for studies of, say, form factors of hadrons at high momentum
transfer.

\subsection{Poincar$\acute{e}$ Generators in Light Cone Field Theory}

   Since the Poincar$\acute{e}$ generators depend on the specific
Lagrangian, as discussed above, let us consider the special
case of the light cone representation of the Poincar$\acute{e}$ 
operators for the scalar field theory, $x^\mu = (x^+,x^1,x^2,x^-)
\equiv (x^+,\vec{x}_\perp,x^-)$ and
$ {\cal L} = \frac{1}{2}(\partial_\mu \phi(x))^2 -\frac{1}{2}\phi(x)^2
 +{\cal L}_{int}(x)$. The field satisfies the commutation rule
 $ [\phi(x),\phi(y)] = \delta(\vec{x}_\perp- \vec{y}_\perp) 
\delta(x^+-y^+) \delta(x^--y^-)$.

The energy momentum tensor has ``good'' operators
\beq
\label{Tgood}
   T^{++} &=& (\partial_+ \phi)^2 \\ \nonumber
   T^{+i} &=& \partial_+\phi\partial_i\phi \; \; i=(1,2) \\ \nonumber
   T^{ij} &=& \partial_i\phi\partial_j\phi,
\eeq
and the ``bad'' operator
\beq
\label{Tbad}
  T^{+-} &=& \frac{1}{2} \partial_+\phi\partial_-\phi - {\cal L}
 \; \; =\frac{1}{2}[\nabla_\perp \phi)^2 + m^2 \phi^2] -{\cal L}_{int}.
\eeq
From this we obtain the seven ``good'' Poincar$\acute{e}$ generators (i=(1,2))
\beq
\label{Pgood}
   P^+ &=& \int d^2\vec{x}_\perp dx^-(\partial_+ \phi(x))^2 \\ \nonumber
   P^i &=& \int d^2\vec{x}_\perp dx^-\partial_+ \phi(x)\partial_i \phi(x) \\ 
\nonumber
   J^3 &=& \int d^2\vec{x}_\perp dx^-[x^1(\partial_+ \phi(x))\partial_1 \phi(x)
  -x^2(\partial_+ \phi(x))\partial_2 \phi(x)] \\ \nonumber
  K^3 &=& \int d^2\vec{x}_\perp dx^-[x^+(\partial_+ \phi(x))^2
  -x^2(\partial_+ \phi(x))\partial_2 \phi(x)] \\ \nonumber
  M^{+i} &=& \int d^2\vec{x}_\perp dx^-[x^+(\partial_+\phi(x))\partial_i\phi(x)
  -x^i(\partial_+ \phi(x))^2].
\eeq
Again we find that the boost in the z direction is interaction-free.
The three ``bad'' Poincar$\acute{e}$ generators, containing the interaction,
are
\beq
\label{Pbad}
  P^- & \equiv & H \; \; =\int d^2\vec{x}_\perp dx^- [\frac{1}{2}
 (\nabla_\perp \phi)^2 + m^2 \phi^2) -{\cal L}_{int}] \\ \nonumber
  M^{-i} &=& \int d^2\vec{x}_\perp dx^-[x^-(\partial_+\phi(x))\partial_i\phi(x)
 -x^i(\frac{1}{2} (\nabla_\perp \phi)^2 + m^2 \phi^2) -{\cal L}_{int})].
\eeq

   This completes our review of light cone field theory. We now review
the light cone formulation of the Bethe-Salpeter (B-S) equation, from which
one obtains the B-S amplitudes needed for all hadronic studies, and the
Schwinger-Dyson (S-D) equation, from which one obtains the dressed quark
propagator needed for the B-S equations for all mesons and baryons.

\section{Bethe-Salpeter in a Light Cone Representation and the Pion Form 
Factor}

   In this section the Bethe-Salpeter (B-S), Schwinger-Dyson(S-D)
formalism is reviewed. Only quark-antiquark systems are considered for
the B-S equation, with application to the pion form factor. We refer to a 
quark-antiquark system as a two-quark system for simplicity. In the
present section we review applications to the pion form factor. In the
next section a recent light cone solution of the quark S-D equation is 
discussed

\subsection{Instant Form of the Bethe-Salpeter Equation}

   The Bethe-Salpeter equation is an exact equation for the  two-particle
propagator. For a derivation see, e.g., Ref\cite{iz}. For a two-quark
propagator the B-S equation is
\beq
\label{B-S}
 (i\not\!{\partial}_{x_1} -m -\Sigma) (i\not\!{\partial}_{x_2} -m -\Sigma) 
 S(x_1,x_2;y_1,y_2)_{2q} &=& S(x_1,y_1)S(x_2,y_2) \\ \nonumber
  + \int d^4 z_1 d^4 z_2 K(x_1x_2;z_1z_2) S(z_1,z_2;y_1,y_2)_{2q},
\eeq
where $S_{2q}$ is the two-quark propagator, $S$ is the quark propagator,
m is the current quark mass, $\Sigma$ is the quark self-energy (from the
S-D equation, shown below), and $K$ is the B-S kernel. 

If the two-particle
system has a bound state with mass M, then there is a pole in $S_{2q}$ in
the total momentum P variable,
\beq
\label{B-Samp}
  S(x_1,x_2;y_1,y_2)_{2q}|_{P^2 \rightarrow M^2} &=& i\frac{\bar{\Psi}(x_1,x_2)
 \Psi(y_1,y_2)}{P^2-M^2}.
\eeq
The quantity $\Psi$, which in the rest system of the bound state of the
two-quark system $|M>$ is
\beq
\label{B-Swf}
 \Psi(x_1,x_2) &=& <0|T[\Psi(x_1)\Psi(x_2)]|M>,
\eeq
which is often referred to as the B-S wave function or B-S amplitude. 
It satisfies the B-S equation
\beq
\label{B-Swfeq}
(i\not\!{\partial}_{x_1} -m -\Sigma) (i\not\!{\partial}_{x_2} -m -\Sigma) 
 \Psi(x_1,x_2) &=& \int d^4 y_1 d^4 y_2 K(x_1x_2;y_1y_2) \Psi(y_1y_2).
\eeq

\begin{figure}
\begin{center}
\epsfig{file=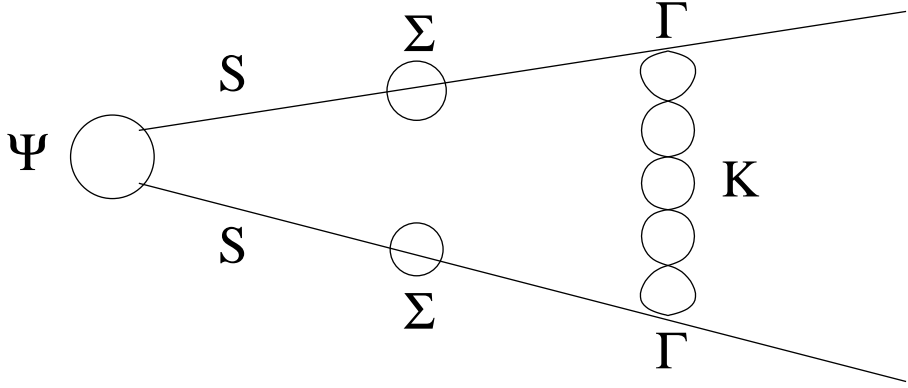,width=12cm}
\caption{Schematic form of B-S amplitude}
{\label{BSamp}}
\end{center}
\end{figure}

   The B-S amplitude and equation is pictured in Fig~\ref{BSamp}.
Note that the solution requires a set of coupled equations for the vertices,
$\Gamma$, self mass of the quark (S-D equation), and the B-S amplitude. 
The Kernal for a perturbative calculation is given by the sum of all 
irreducible diagrams. For nonperturbative theories like QCD one must use 
models.

\subsubsection{Problems With Instant Form of B-S Equation}

   About a half century ago an important paper by Wick\cite{wick54} pointed 
out a number of serious problems with the B-S equation and B-S amplitude.
One of the problems is the dependence on the relative time (or relative energy)
variable, whose physical meaning never was clear. He found a solution to
this problem via the Wick rotation to Euclidean space. This has
wide application, and is still used in hadronic physics. Since this problem
has been solved by the light cone formulation we do not review this idea,
but aspects will appear later in this review.

\subsection{Light Cone Bethe-Salpeter Equation}

  As discussed above, in order to obtain a B-S wave function for hadronic
physics one must also solve the Schwinger-Dyson equation for the dressed
quark propagator, which we discuss in the following section. In any case,
models are needed for QCD. Here we discuss some early models of the light
cone formulation of the B-S equation and applications to the pion form
factor.

  Since in the infinite momentum frame the pair creation and annihilation
processes are suppressed\cite{wein66}, it has many of the attributes of the
light cone formulation. In an early paper\cite{chma69} QED at infinite
momentum was treated, and this work has been used recently for obtaining
a light cone S-D equation, discussed in the next section. The light cone
form of the B-S equation can be obtained by using the $\infty$-momentum 
frame\cite{saw85} or projection onto the light cone\cite{bjs85}. In 
momentum space, using the notation for the momenta of the two quarks
$k_1 = (x_1P^+,\vec{k}_\perp,k_1^-), k_1 = (x_2P^+,-\vec{k}_\perp,k_2^-)$,
with $k_i^- = (m^2 +k_\perp^2)/k_i^+$, the light cone equation for the
B-S wave function, $\psi(k_1,k_2,P)$ is
\beq
\label{lcBS}
 (M^2 -\frac{m^2+k_\perp^2}{x_1x_2})\Psi(x_1,x_2,\vec{k}_\perp) &=& 
  \int_{0}^{1}dy_1dy_2\delta(1-y_1-y_2)d^2l_\perp \\ \nonumber
 && K(x_1,x_2,\vec{k}_\perp,y_1,y_2,\vec{l}_\perp)\Psi(y_1,y_2,\vec{l}_\perp).
\eeq
In this form one assumes that the mass, m, of the quark is a constant
constituent quark mass, and uses a model for the kernel.

\subsection{Application of Light Cone B-S Amplitudes to Pion Form Factor}

   In this subsection we consider applications of the light cone formalism
 to the pion form factor. The importance of the light cone formalism to
study the transition of the pion form factor from a low-momentum region
in which nonperturbative effects dominate is first reviewed, followed by
treatment with a fixed quark mass, and then the most recent work with a
running quark mass. See Sections 8 and 9 for details on l-c wave functions.

\subsubsection{Light Cone Wave Function and Transition to Asymptotic QCD}

   One of the interesting problems in the past two decades is the nature 
of the transition from the low-momentum region of nonperturbative QCD to
the asymptotic region of perturbative QCD. This problem has been studied
theoretically for many years, such as the modified quark distribution 
functions\cite{cz82}, and experimentally. An early application of light cone
methods\cite{jk86,kw93} was simply the use of a light cone vs instant form 
pionic wave functions. The result is quite startling and a dramatic
illustration of the advantage of a light cone representation.

\begin{figure}
\begin{center}
\epsfig{file=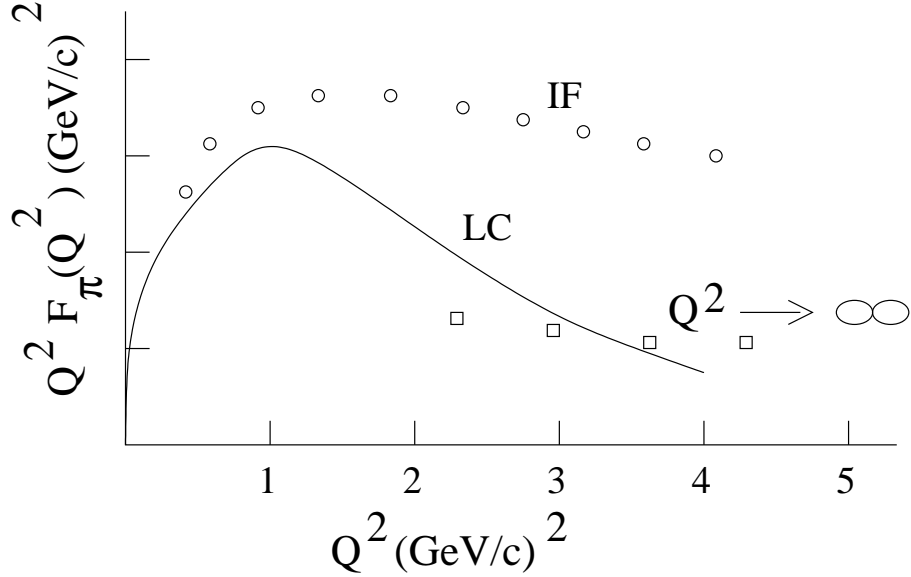,width=12cm}
\caption{Pion form factor in a light cone representation (LC, solid) and with
an instant form (IF, dotted) harmonic oscillator wave function}
\label{pionffho}
\end{center}
\end{figure}

   Given the wave function of the pion, $\phi_\pi$, for the low-momentum
or soft part of the pion form factor with a nonrelativistic wave function 
the pion form factor is simply
\beq
\label{fpinr}
      F_\pi^{soft}(Q^2) &=& \int d^3 p \phi_\pi^*(\vec{p},\vec{P}+\vec{Q})
 \phi_\pi(\vec{p},\vec{P}).
\eeq
On the other hand it is known that the high momentum transfer (hard) limit 
perturbation diagrams can be used, with the result\cite{fj79}
\beq
\label{farrar}
    F_\pi(Q^2)|_{Q^2 \rightarrow \infty} & \simeq & 8\pi f_\pi^2
\frac{\alpha_s(Q^2)}{Q^2},
\eeq
with a similar result obtained by others\cite{lb80}. It is of great interest
to know when perturbation theory can be applied, which is when the
nonperturbative soft contributions become smaller than the high-Q 
contributions.  In an attempt to study this a nonrelativistic harmonic 
oscillator wave function
\beq
\label{ho}
 \phi_\pi(\vec{p}) &=& \frac{1}{[\pi b]^{3/4}}e^{-(p_z^2 +p_\perp^2)}  
\eeq
 was used\cite{ils84}. Even at $Q^2 \simeq$ 10 GeV$^2$ the
soft part dominates the hard term. On the other hand, recognizing that even
at 1 GeV$^2$ relativistic effects are important, a calculation with the
the same wave function in the light front representation\cite{jk86}:
\beq
\label{lcho}
    \phi_\pi(\vec{p},\vec{P}) &=& N e^{-\frac{P^2 +p_\perp^2}{2b^2}} 
 e^{-\frac{x^2}{4 b^2} \frac{p_\perp^2+m^2}{1-x^2}},
\eeq
with N a normalization constant, has been carried out.
The results are shown in Fig~\ref{pionffho}.
As one can see by comparing the light cone (LC) with the instant form (IF)
curves, at Q$^2$ greater than about 0.5 GeV$^2$ the relativistic corrections
become increasingly important. The curve labeled $Q^2 \rightarrow \infty$
gives the limit of the hard scattering process\cite{fj79}. In the light cone
model the soft part of the form factor becomes smaller than the asymptotic
hard part for $Q^2 \simeq 4 GeV^2$, while in the nonrelativistic instant 
form\cite{ils84} the soft part dominates until very high Q$^2$. A vastly
different interpretation of the transition from soft to hard QCD results.

\subsubsection{Light Cone B-S Amplitude for Pion Form Factor}

   There have been a number of studies of the soft (nonperturbative)
pion elastic form factor using light cone or other relativistic models
\cite{dz88,car96} and an early study\cite{dm87} using light cone wave 
functions to test the amplitudes of Ref.\cite{cz82}, as well as the
work of Ref.\cite{jk86}.

   In this subsection we discuss the
study of the B-S equation for the pion amplitude\cite{jk90} with a
calculation of the pion form factor including both the soft and hard parts.
The starting point is the light cone B-S equation, Eq.(\ref{lcBS}), using 
the kernel $K = K_1 + K_2$, with $K_2$ the one gluon
exchange kernel used to get the hard asymptotic form for the pion form
factor\cite{fj79,lb80}, and $K_1$ modeled to give the soft confining part:
\beq
  K(x_1,x_2,\vec{k}_\perp,y_1,y_2,\vec{l}_\perp)  &=& 
-\frac{b m^2 \Theta(y_1-x_1)}{x_1x_2(y_1-y_2)^2}[M^2-\frac{k_\perp^2+m^2}
{x_1}-\frac{l_\perp^2+m^2}{y_2}  \\ \nonumber
  &&  -\frac{(\vec{k}_\perp-\vec{l}_\perp)^2}{(y_1-x_1)^2}]^{-2}
 + (1 \leftrightarrow 2),
\eeq
with b an interaction strength, and m is a mass parameter.
The $\pi^+$ form factor is given by
\beq
\label{fpilc}
   F_\pi &=&\int \frac{dx}{1-x^2} \frac{d^2\vec{k}_\perp}{16\pi^3} 
[ \frac{2}{3}\Psi^{\dagger}(\vec{k}_\perp +\frac{1-x}{2}\vec{q},x)
 + \frac{1}{3}\Psi^{\dagger}(\vec{k}_\perp  -\frac{1-x}{2}\vec{q},x)]
\Psi(\vec{k}_\perp,x). 
\eeq
The B-S amplitude is constrained by 1) normalization, 2) the pion
decay constant, $f_\pi$, and 3) the pion mass. The pion mass is quite
difficult to fit with this model and the best fit to the other
parameters resulted in a pion mass of 356 MeV. To obtain a solution,
the spinors are first projected out. Expressing the B-S amplitude as
$\Psi(x_1,x_2,\vec{k}_\perp) = \Phi(x_1,x_2,\vec{k}_\perp)
V(\vec{p}_1)V(\vec{p}_2)X_F$, where $V(\vec{p})$ is the melosh 
operator\cite{mel74,kt80} and $X_F$ are the quark light cone spinors.
After projecting out the spinors, the B-S equation for 
$\Phi(x_1,x_2,\vec{k}_\perp)$ is solved using the technique of expansion
in hyperspherical harmonics\cite{saw85}.

\begin{figure}
\begin{center}
\epsfig{file=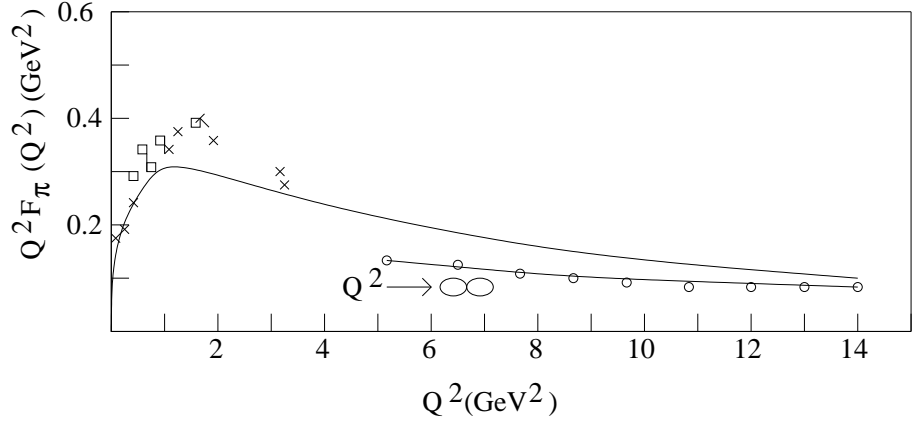,width=12cm}
\caption{Pion form factor in from B-S amplitude calculated with a light
cone kernal containing confining and asymptotic parts}
\label{pilcff}
\end{center}
\end{figure}

   The result for the pion form factor is shown in Fig.~\ref{pilcff}.
The prediction is that the asymptotic part of the B-S amplitude dominates
by 10-15 GeV$^2$. Other light cone model calculations\cite{kw93} gave
similar results.

\subsubsection{The Running Quark Mass and the Pion Form Factor}

   In addition to the transition from soft to hard processes dominating
hadronic form factors, as one goes from the nonperturbative to the
perturbative regions of momentum transfer the effective quark mass also 
undergoes an evolution from the constituent mass of low-energy quark
models to the current quark mass in the QCD Lagrangian. This can be
understood most directly by considering the quark propagator, which
has the form
\beq
\label{qprop}
  S(p)^{-1} &=&  A(p^2)\not\!p-B(p^2),
\eeq
where the running quark mass is given by $M(p^2) = B(p^2)/A(p^2)$.
The functions $A(p^2), B(p^2)$ are determined by a set of self-consistent
equations in the Schwinger Dyson formalism, which we review in the 
following section.  $M(p^2)$ is expected to undergo its transition from 
the constituent mass of about 300 MeV to a few Mev for u and d quarks
at about 1 GeV.

   The effect of this running quark mass on the pion form factor was
studied some years ago\cite{kw94}, and in greater detail recently\cite{kcj01}.
The pion form factor, given by Eq.(\ref{fpilc}) and the discussion following
that equation, is modified if one includes the dressed quark vertex.
The elastic pion form factor can be written as
\beq
\label{pikcj}
   F_\pi(Q^2) &=& \frac{-1}{Q^2} \int^1_0 dx\int d^2{\bf k}_\perp
\Psi^{*}_{\lambda'_q\lambda{\bar q}}(x,{\bf k'}_\perp)\nonumber\\
&\times&\frac{{\bar u}_{\lambda'_q}(p'_q)}{\sqrt{p^{'+}_q}}\Gamma^\mu
\frac{u_{\lambda_q}(p_q)}{\sqrt{p^+_q}}
\Psi_{\lambda_q\lambda{\bar q}}(x,{\bf k}_\perp),
\eeq
where $p^+_q$=$p^{'+}_q$=$(1-x)P^+$ and 
${\bf k'}_\perp$=${\bf k}_\perp -x{\bf q}_\perp$ in the initial pion
rest frame, ${\bf P}_\perp$=0. The helicity of the quark(antiquark) is 
denoted as $\lambda_{q({\bar q})}$.  The B-S amplitude is the one used 
in Ref.\cite{jk90}
\beq
\label{kcjwf}
\Psi_{\lambda_q\lambda_{\bar q}}(x,{\bf k}_\perp) &=&
 \sqrt{\frac{\partial k_z}{\partial x}}\Phi(x,{\bf k}_\perp)
{\cal R}_{\lambda_q\lambda_{\bar q}}(x,{\bf k}_\perp),
\eeq
where $\Phi$, defined in Eq.(\ref{lcho}) is the radial wave function
and $\sqrt{\frac{\partial k_z}{\partial x}} {\cal R}$ is the spinor wave 
function. The Ball-Chiu\cite{bc80}
ansatz is used for the vertex (see the discussion in the next section)
\beq
\label{BC}
\Gamma^\mu_{\rm BC} &=& \frac{ (\not\!p + \not\!p')}{2}(p + p')^\mu
\frac{ A(p^{'2})-A(p^2)}{p^{'2}-p^2} \nonumber\\
&+& \frac{A(p^{'2})+A(p^2)}{2}\gamma^\mu
- (p + p')^\mu\frac{B(p^{'2})-B(p^2)}{p^{'2}-p^2}.
\eeq

\begin{figure}
\begin{center}
\epsfig{file=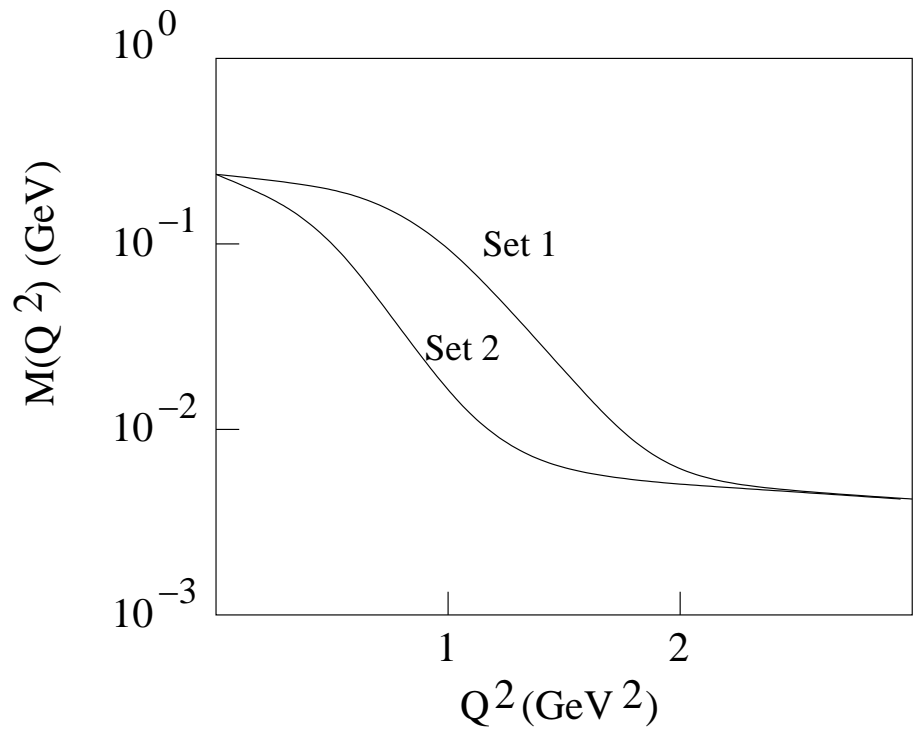,width=12cm}
\caption{The Running Quark Mass. Set 1, Set 2 defined in text}
\label{masskcj}
\end{center}
\end{figure}

   In order to consistently determine the running quark mass, the S-D
equation must be solved in a light cone representation, which is reviewed
in the next section. In Ref.\cite{kcj01} models were used. 
For the crossing antisymmetric (CA) parameterization the model is
\beq
\label{ME2}
M(p^2) &=& m_0 + (m_c - m_0)\frac{ 1+\exp(-\mu^4/\lambda^4)}
{1 +\exp[(p^4 - \mu^4)/\lambda^4]},
\eeq
where $m_0$ = 5 MeV and $m_c$ = 220 MeV are the current and constituent 
quark masses, respectively. 
The parameters $\mu$ and $\lambda$ are used to adjust the
shape of the mass evolution. For the CA picture the two sets of parameters
used are $(\mu^2,\lambda^2)$ = (0.95,0.63) [Set 1] and (0.28,0.55) [Set 2]
GeV$^2$. The resulting $M(Q^2)$, with $Q^2 = -p^2$, is shown in 
Fig.~\ref{masskcj}.

\begin{figure}
\begin{center}
\epsfig{file=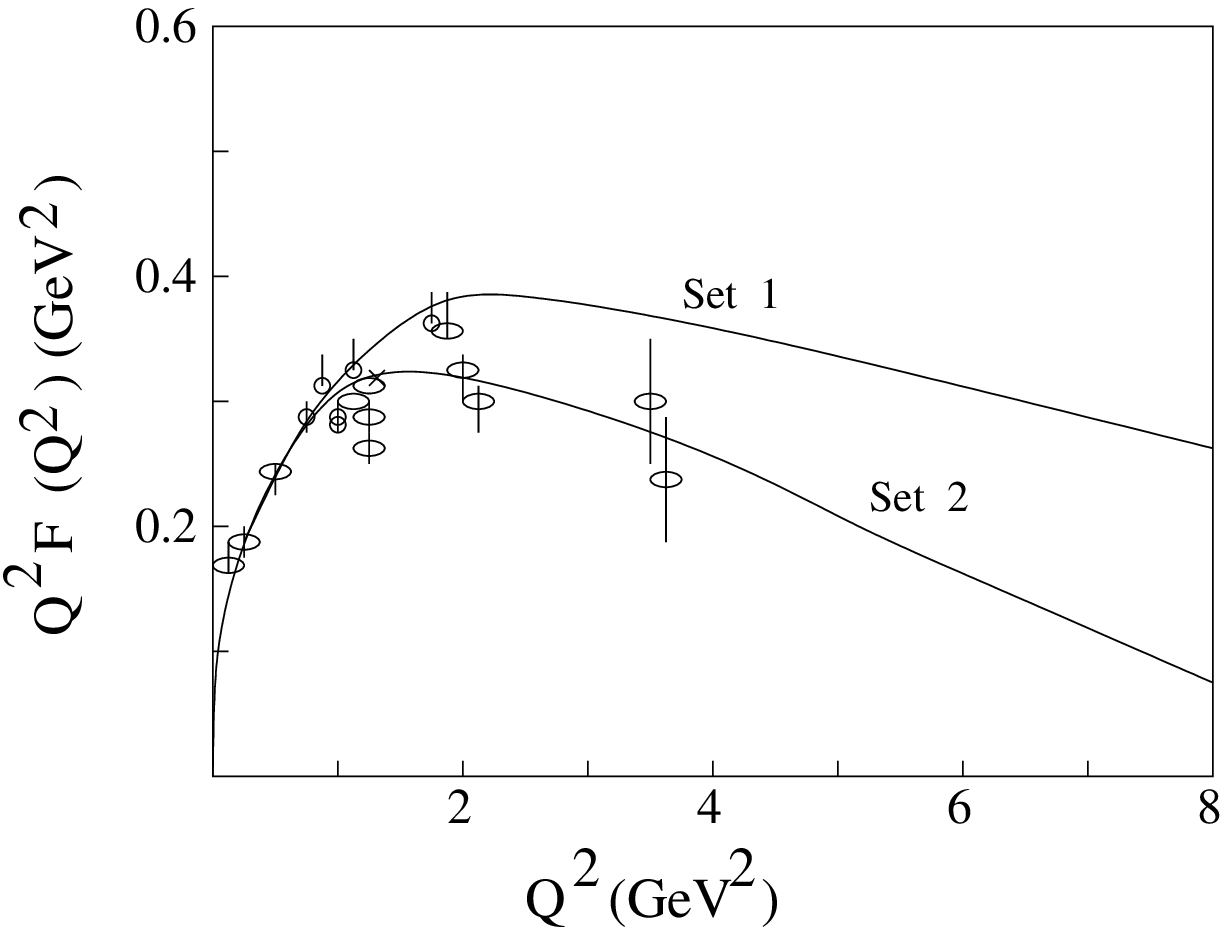,width=12cm}
\caption{Pion Form Factor From a soft B-S Amplitude With An Evolving
Quark Mass}
\label{piffkcj}
\end{center}
\end{figure}

   The results of this model for the pion form factor are shown in 
Fig.~\ref{piffkcj}.
 In comparison with Fig.~\ref{pilcff} it is
clear that the running quark mass must be taken into consideration
in order to trace the evolution from the soft perturbative region
to the asymptotic region.

\subsubsection{Other Light Cone Calculations of the Pion Form Factor}

   There has not been much experimental progress in extending accurate
measurements of the pion elastic form factor to higher momentum transfers,
and there is little hope in that the region above 10 GeV$^2$ will be
reached soon. There have been suggestions for improving the perturbative
calculations, such as Refs.\cite{bjp,sjr,wk96}, but detailed calculations of
the pion form factor from the soft to the hard regions of momentum
transfer require models of the nonperturbative part, and little progress
has been made over the past decade. This is unfortunate, since the soft
to the hard transition for the pion occurs at a much lower momentum transfer
than for the nucleon, but the experiment is quite difficult.

\subsection{Low Energy Theorems, Chiral Symmetry, and Light Cone Field Theory}

   Although light cone field theory is generally most important at high
momentum transfers, due to the difficulty in carrying out Lorentz 
transformations in the instant form (as discussed above), there has been at
least one important application at very low energies. We briefly review
the soft pion theorems for threshold pion photoproduction and the importance
of a light cone formulation. A complete discussion with references can be
found in Ref.~\cite{ck90}.

   The amplitude for pion photoproduction with a nucleon target,
$N(\gamma,\pi)N$ has three isospin amplitudes, $A^{(+)}, A^{(-)}, A^{(0)}$.
A model-independent expansion in the low energy parameter 
$\mu \equiv m_\pi/M_N$ gives\cite{kr54}
\beq
\label{sp1}
     A^{(-)} &=& 1 + O(\mu^2) \nonumber \\ 
     A^{(+)} &=& -\mu/2 + 0(\mu^2) \nonumber \\
     A^{(0)} &=& -\mu/2 + 0(\mu^2) \; ,
\eeq
or in the soft pion limit 
\beq
\label{sp2}
                      A^{(-)} &=& 1. \nonumber \\ 
                      A^{(+)} &=&  A^{(0)} \; .
\eeq

   Standard quark models have a great difficulty with these model-independent
predictions. The main problem is that the pion photoproduction processes
are related to pion exchange currents, and standard quark models give
very different predictions at low energy than hadronic models\cite{bdg83}.
As discussed in Ref.~\cite{ck90}, the low-energy limit in standard quark
models gives
\beq
\label{sp3}
             A^{(+)} &=& \frac{5}{9}  A^{(0)} \; .
\eeq
This is a serious violation of low energy theorems, which is also found in
chiral quark models\cite{bdg83}. 

   This problem is resolved in light cone field theory models. An interesting
observation is that for the calculation of $ A^{(-)}$ the standard pion 
exchangs process is replaced by the instantaneous term\cite{lb80}. It was 
shown in \cite{ck90} that with a standard light cone harmonic oscillator model
\beq
\label{sp4}
  \phi (x,\vec{k}_\perp) &\sim& e^{\left(-\frac{k_\perp^2 +m^2}{b^2 x}\right)}
\eeq
the low-energy theorem predictions of Eq.(\ref{sp1}) are satisfied. 
It should also be noted that the calulation of the $E_{o+}$ amplitude
does not agree with low-energy theorems, and that the measurement of the
$p(\gamma,\pi^0)p$ threshold cross section\cite{beck90} obtained a result
for the $E_{o+}$ amplitude also in disagreement with low-energy theorems.

\section{Light Cone Representation of the Quark Schwinger-Dyson Equation}

  In this section we discuss the light cone formulation the quark
Schwinger-Dyson Equation (SDE). There has been a great deal of interest
in the SDE for hadronic physics, with excellent reviews\cite{rw94,
tandy97}. The Schwinger-Dyson (S-D) formalism is a technique for 
obtaining the dressed quark propagator; although this propagator 
is not in itself physical, it can
be used to obtain nonperturbative hadronic strucdture. We review this in
the next section. The quark propagator in space-time is defined as the 
correlator
\beq
\label{quarkprop}
    S_{\mu\nu}^{ab} &=&  <0|T[q_\mu^a(x)\bar{q}_\nu^b(0)]|0>,
\eeq
where T is the time-ordering operator.
Since the propagator is diagonal in color, the propagator can be
written as $ S(x)= <0|T[q^a(x)\bar{q}^a(0)]|0>$, with no sum on the color
index a. The S-D formalism starts with an exact expression for the quark
propagator and explores models to obtain solutions in the presence 
of nonperturbative effects.

\subsection{Schwinger-Dyson Formalism}
\begin{figure}
\vspace*{15 cm}
\includegraphics{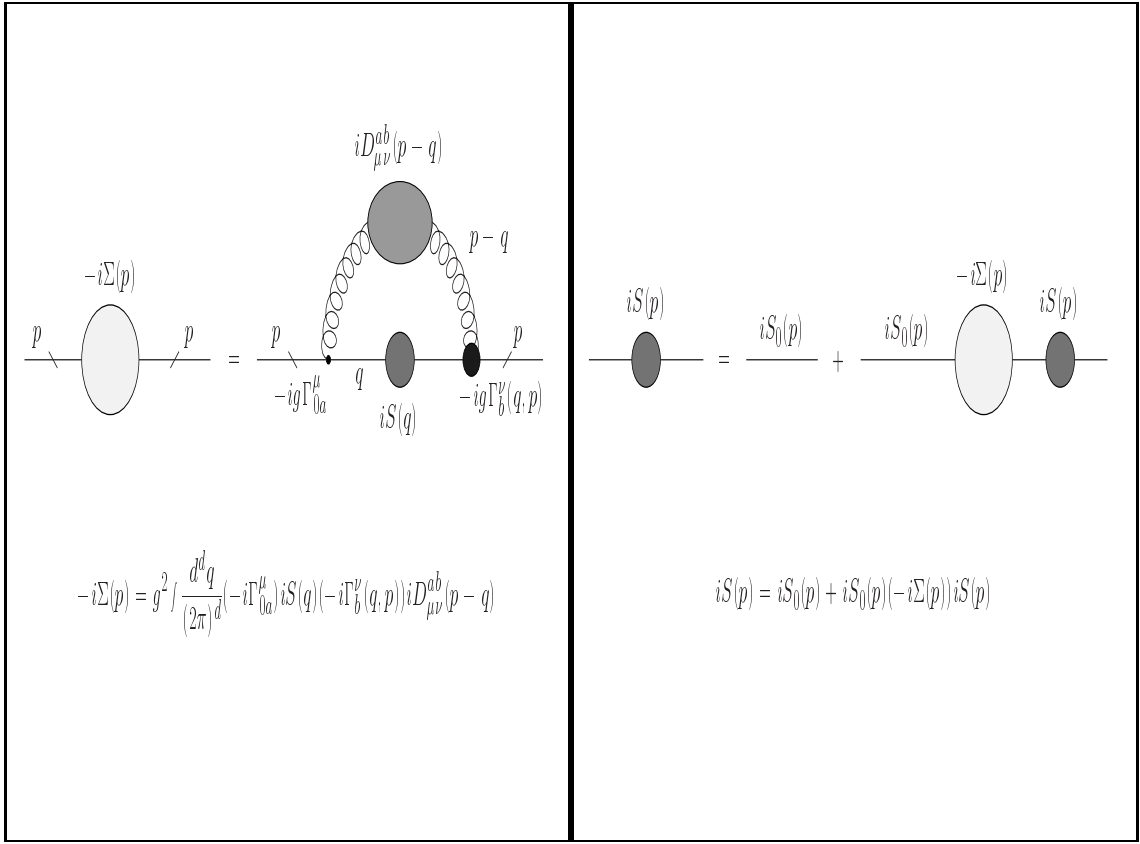}
\caption{Diagrammatic representation of the SDE for the quark propagator.} 
\label{sdkl}
\end{figure}

   The full dressed quark propagator satisfies the 
Schwinger-Dyson equation (SDE). (See Fig~\ref{sdkl}):

\beq
\label{s-deq}
  S^{-1}(p) & = & S_0^{-1}(p)-\Sigma(S(p)) \nonumber \\
       \Sigma(S(p)) &=& ig^2\int \frac{d^dq}{(2\pi)^d}
\gamma^{\mu} \frac{\lambda_a}{2}S(q)\Gamma^{\nu}_{b}(q,p)
D^{ab}_{\mu \nu}(p-q)
\eeq
where $\Gamma^{\nu}_{b}(q,p)$ and $D^{ab}_{\mu \nu}(p-q)$ are the dressed
quark-gluon vertex and dressed gluon propagator, $\lambda_a$ is a SU(3)
operator in color space, $\gamma_\mu$ is a Dirac operator, with 
Greek letters representing Lorentz indices and latin letters standing
for color indices. The SDE is illustrated in Fig.\ref{sdkl}
\begin{figure}
\begin{center}
\epsfig{file=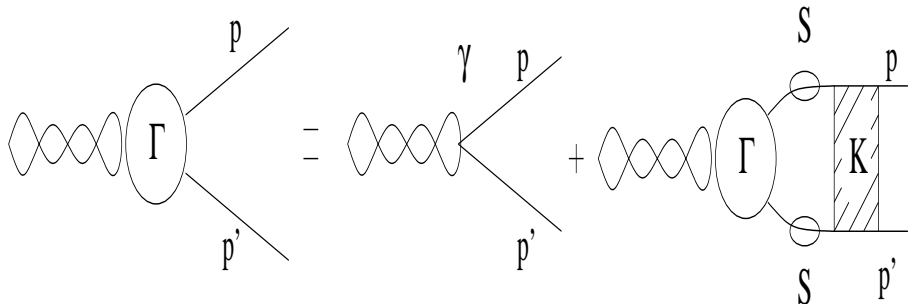,width = 12 cm,height=4cm}
\caption{Diagrammatic representation of the SDE for the dressed
gluon-quark vertex.} 
\label{sdvertex}
\end{center}
\end{figure}

  The renormalized vertex is obtained from the vertex SDE
\beq
\label{sdver}
  \Gamma^{\nu}(p,p') &=&  Z\gamma^\nu +\sum_{b=1}^{n_c}
\int \frac{d^dq}{(2\pi)^d}S^b(p'+q)\Gamma^{\nu}_{b}(p+q,p'+q)S^b(p+q)
 K^b(p',p,q),
\eeq
where Z is a renormalization constant and the kernel $K(p',p,q)$ is 
related to the dressed gluon propagator $D(q)$. This is illustrated in
Fig.~\ref{sdvertex}

The solutions of the SDE are of the form given in Eq.(\ref{qprop})
\beq
\label{qprop1}
  S(p)^{-1} &=&  A(p^2)\not\!p-B(p^2). \nonumber
\eeq
with the effective mass given by  $M(p^2) = B(p^2)/A(p^2)$. 

The necessary ingredients for obtaining solutions are the dressed vertex 
and kernel. In practice, the dressed vertex is modeled using symmetries, 
and the main result of S-D calculations for applications to hadronic 
physics, in addition to the dressed quark propagator itself, is the dressed 
Gluon kernel. The transverse part of the dressed vertex is constrained by
the Ward-Takahashi identity (see Ref\cite{iz} for detailed discussions)
\beq
\label{wti}
   (p' - p)_\mu \Gamma^{\nu}(p,p') &=& S^{-1}(p') -S^{-1}(p)
\; \; \; \; {\rm Ward-Takahashi \; Identity},
\eeq
which is an extension to nonvanishing momentum transfer of the Ward identity,
\beq
\label{wardi}
    \Gamma^{\nu}(p,p') &=& \frac{\partial}{\partial p_\mu} \Sigma(p)
\; \; \; \; {\rm Ward \; Identity}.
\eeq
This leads to the Ball-Chiu ansatz, Eq.(\ref{BC}), and the Curtis-Pennington
forms\cite{cp90}, discussed in detail in Ref.\cite{rw94}.

   The general form of the gluon propagator is
\begin{equation}\label{gluongprop}
D_{\mu \nu}(k) = \frac{-1}{k^2+i\epsilon}\left\{
\left( g_{\mu \nu}- \frac{ k_{\mu}k_{\nu} }{k^2+i\epsilon}\right)D(k^2)
+\xi \frac{ k_{\mu}k_{\nu} }{k^2+i\epsilon}\right\}
\end{equation}
where $\xi$ is the gauge parameter. The  Landau gauge corresponds to the
choice of $\xi = 0 $, while a so-called Feynman-like gauge corresponds to
the choice
\beq
\label{feynmanlike}
    D_{\mu \nu}^{a b}(x) &=& \delta_{ab} \delta_{\mu \nu}D(x)
\eeq
for the model gluon propagator.

\subsubsection{Constraints on Quark Propagator}

\begin{figure}[ht]
\begin{center}
\epsfig{file=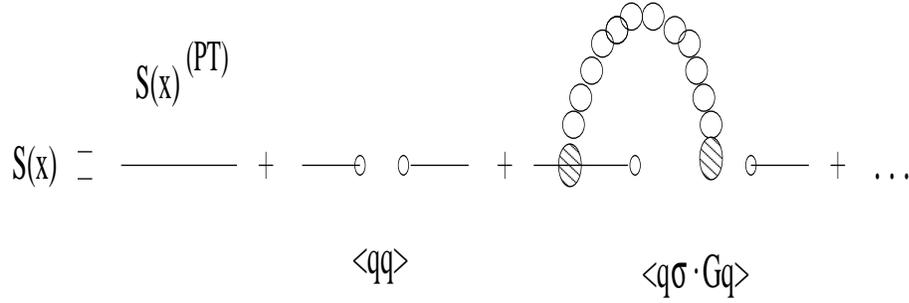,width = 12 cm,height = 4 cm}
\caption{Diagrammatic representation of the operator product expansion for 
the quark propagator.} 
\label{ope}
\end{center}
\end{figure}

   One approach to the treatment of the nonperturative aspects of QCD
correlators is to use the operator product expansion, which implies
a continuation to large momenta. The operator product expansion of the
quark propagator is
\beq
\label{qpropope}
    <0|T[q_\mu^a(x)\bar{q}_\nu^b(0)]|0> &=&  S_{\mu\nu}^{ab (PT)}
  +\delta_{ab}\delta_{\mu\nu}[-\frac{<\bar{q}q>}{12} + \frac{x^2 g_s
<\bar{q}\sigma\cdot G q>}{192}] +...,
\eeq
where $<\bar{q}q>$ is the quark condensate and $<\bar{q}\sigma\cdot G q>
= <\bar{q}\sigma_{\mu\nu} G^{\mu\nu} q>$ is the mixed condensate. This is
illustrated in Fig.~\ref{ope}.
The condensates are vacuum expectation values of local operators, and 
therfore are gauge invariant constants. The quark condensate has been known
for many decades, and the mixed constant is somewhat uncertain, but has
been estimated using the QCD sum rule method. These condensates can
be used to constrain the S-D solutions by the relationships in Euclidean
space
\beq
\label{condensates}
   <0|:\bar{q}(0) q(0):|0> &=& 
-\frac{3}{4\pi^2} \int_0^{+\infty} dS S \frac{B(S)}{SA^2(S) + B^2(S)}
 \\
  <0|:\bar{q}(0)g \sigma \cdot G(0) q(0):|0> &=&
  \frac{9}{4 \pi^2} \int_0^{+\infty} dS S \{ 
 \frac{81 B(S)[2SA(S)(A(S)-1) + B^2(S)]}{16(SA^2(S) + B^2(S))} \nonumber \\
       && + S\frac{B(S)(2-A(S))}{SA^2(S) + B^2(S)} \} \; , \nonumber 
\eeq
with the notation that $S=Q^2$ the Euclidean squared momentum.
Another constraint is given by
\beq
\label{pidecay}
   f^2_{\pi} &=& \frac{3}{4\pi ^2}\int_0^{+\infty}dS \frac{S M(S)}{A(S) 
[S+M^2(S)]^2} \left[M(S)-\frac{S}{2}\frac{dM}{dS}\right],
\eeq
with M=B/A.

\subsubsection{Rainbow Approximation For SDE}

   The rainbow approximation corresponds to the choice of 
\beq
\label{rainbow}
   \Gamma^{\mu}_{a} &=& \gamma^{\mu} \lambda_a/2 \;\;{\rm rainbow \; 
approximation}).
\eeq
This is the main feature of the Global color model\cite{cr85}, reviewed in
Ref.\cite{tandy97}. Using the Feynman-like gluon propagator of 
Eq.(\ref{feynmanlike}) the self mass is given as
\beq
\label{gcmass}
    \Sigma(p) &=& \frac{4}{3}g_s^2\int \frac{d^4q}{(2\pi)^4}
\gamma^{\mu} S(q)\gamma^{\nu}_{b}(q,p) D(p-q).
\eeq
In this rainbow approximation the quark SDE becomes a set of
coupled integral equations
\beq
\label{ABeqns}
   [A(p^2)-1] &=& g_s^2\frac{8}{3}\int \frac{d^4q}{(2\pi)^4}D(p-q)
 \frac{A(q^2) q\cdot p}{q^2A^2(q^2)+B(q^2)} \nonumber \\
  B(p^2) &=& g_s^2\frac{16}{3}\int \frac{d^4q}{(2\pi)^4}D(p-q)
 \frac{B(q^2)}{q^2A^2(q^2)+B(q^2)}.
\eeq
There have been many applications of this model. The vertex for the 
SDE, Eq.(\ref{sdver}) without the perturbative term has the form
of a B-S amplitude for a bound quark-antiquark state, i.e., a meson. 
The B-S equation for a meson as a bound state with momentum P is 
\beq
\label{SDBS}
  \Gamma^{ab}_{M}(p,P) &=& \int \frac{d^dq}{(2\pi)^4}S^a(q+xP)
\Gamma^{ab}_{M}(q,P)S^b(q-(1-x)P) K(p,q,P),
\eeq
where a,b are the color indices of the quark, antiquark, x is the 
momentum fraction of the quark, and K is the kernel. If one is
given solutions to the quark SDE this provides a model of
mesonic properties (see, e.g., \cite{mr97,bur97,mt99,mt00}).
As discussed in the previous section, one of the most important
studies of hadronic properties is the transition from the soft 
nonperturbative to the hard perturbative region of momentum
transfer, which is best done in a light cone representation.
For this reason it would be most valuable to have a light cone
S-D solution to use to get hadronic B-S amplitudes, analogous to
Eq.(\ref{SDBS}). We now discuss the first successful light cone
solutions of the quark SDE.

\subsection{SDE and Quark Propagator in a Light Cone Representation}

The method used for solving
 Eq.~(\ref{s-deq}) in a light cone representation in Ref.\cite{kl02} is
to use a technique introduced originally for perturbation 
theory\cite{changma69}.  Chang and Ma showed how the rules of 
light cone perturbation theory [LCPT] can be derived in the infinite 
momentum frame from the usual covariant Feynman rules simply by changing 
into light cone variables:
\[ q^{\pm}=q^0 \pm q^{3}, \;\; {\bf q}_{\perp}=
\left( q^1,q^2 \right) \]
An important feature of the new rules is that the range of integration 
over the $q^+$ variable becomes finite 
\[ \int_{-\infty}^{+\infty}dq^+ \longrightarrow \int_0^{p^+}dq^+, \]
where $p^+$ is the ``plus'' (longitudinal) component of the external 
momentum. This feature is related 
to the fact that in LCPT there are no diagrams with lines going 
backwards in time and no vacuum diagrams (with the exception of 
zero-modes and instantaneous terms in fermion propagators). This is 
a crucial property of light cone field theory, since it simplifies 
tremendously the structure of the vacuum. 

An important consideration is the form of the integrals needed for the 
calculation of the quark self-energy $\Sigma$ in the SDE. We are faced with 
the non-perturbative self-energy diagram seen in Fig.~\ref{sdkl}. 
The integrals occuring in the SDE diagram can be written as a sum of terms 
of the form:
\beq
\label{genericintegral}
\int \frac{d^dq}{(2\pi)^d} q^{\mu_1}q^{\mu_2}\ldots 
q^{\mu_n}f_Q\left(q^2+i\epsilon\right)f_G\left((p-q)^2+i\epsilon \right),
\eeq
where the $f_Q$ contains factors coming from the quark propagator and 
possibly from the dressed vertex function, $f_G$ contains factors coming 
from the gluon propagator and also possibly from the dressed vertex function, 
and the factors of $q^{\mu_i}$ come from Dirac operators.
 Since the external momentum $p$ is held fixed during 
this integration, possible factors of $p^{\mu_i}$ and possible dependence 
of the Green's functions on $p^2$ are not relevant to the analysis below. 
Consider first the scalar case, $n=0$ in~(\ref{genericintegral}), i.e., 
no factors $q^{\mu}$. Using the variables:
\[ \alpha=q^+/p^+, \;\; s'=q \cdot q, \;\; s=p \cdot p, \;\; 
{\bf q'}_{\perp}={\bf q}_{\perp}-\alpha \; {\bf p}_{\perp}, \] 
the integral in (\ref{genericintegral}) becomes:
\beq
\label{inmyvariables}
\int \frac{ds'd\alpha d^{d-2}{\bf q'}_{\perp}}{2 | \alpha |(2 \pi)^d}
f_Q\left(s'+i\epsilon\right) 
f_G\left(-\alpha^{-1} {\mathcal P}(q^{'2}_{\perp},s,s',\alpha,\epsilon)\right),
\eeq
where all the integrals are over the entire real line, and
\[ {\mathcal P}(q^{'2}_{\perp},s,s',\alpha,\epsilon) \equiv
q^{'2}_{\perp}+\alpha (1-\alpha)(-s)+(1-\alpha) s'-i \alpha \epsilon \].

   In Ref.\cite{kl02} it was shown that {\em if all the singularities of 
the functions $f_Q$ and $f_G$ occur 
on or below the real axis, only the interval $(0,1)$ in the integration 
over $\alpha$ contributes to the integral}. The model must ensure
that the integrals are ultravioletly convergent and thus closing the contour 
with a semicircle at infinity introduces no additional contribution.
 The integral in (\ref{inmyvariables}) then becomes
\beq
\label{lcform}
\frac{1}{4\pi}
\int_{-\infty}^{+\infty}\frac{ds'}{2\pi} \!
\int_0^1 \! d \alpha \!
\int \frac{d^{d-2}{\bf q'}_{\perp}}{(2 \pi)^{d-2}}
\alpha^{-1}f_Q\left(s'+i\epsilon\right) 
f_G\left(-\alpha^{-1}  {\mathcal P}(q^{'2}_{\perp},s,s',\alpha,\epsilon)
\right).
\eeq
It should be noted that to preserve important features of light cone 
theory (here reflected in the integral over $\alpha$ being over a 
finite range) one 
must make assumptions about the location of the singularities 
in $f_Q$ and $f_G$. These assumptions are much like the ones necessary to 
justify a Wick rotation, however, the method of Ref.\cite{kl02}
allows one to solve the quark SDE for time-like values of the momentum $p$,
which is difficult to do after a Wick rotation.

\begin{figure}[ht]
\vspace*{7.5 cm}
\includegraphics{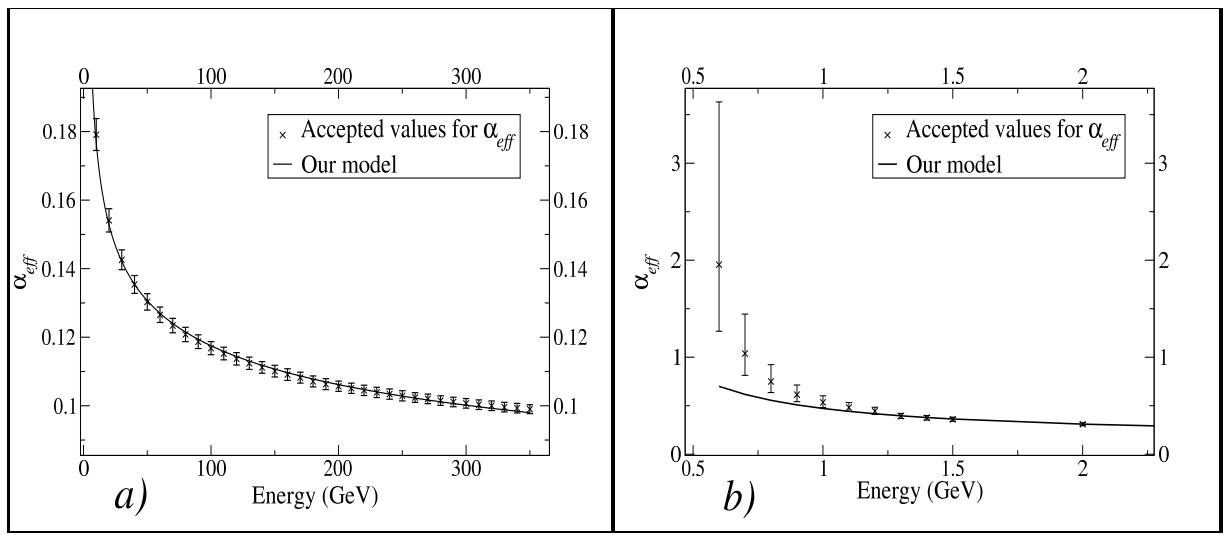}
\caption{Fits to $\alpha_{eff}$ with model of Ref.\cite{kl02} a) for 
energies up to 350 GeV and b) a close-up for lower energies.} 
\label{alphafit}
\end{figure}

\begin{figure}[ht]
\vspace*{7.5cm}
\includegraphics{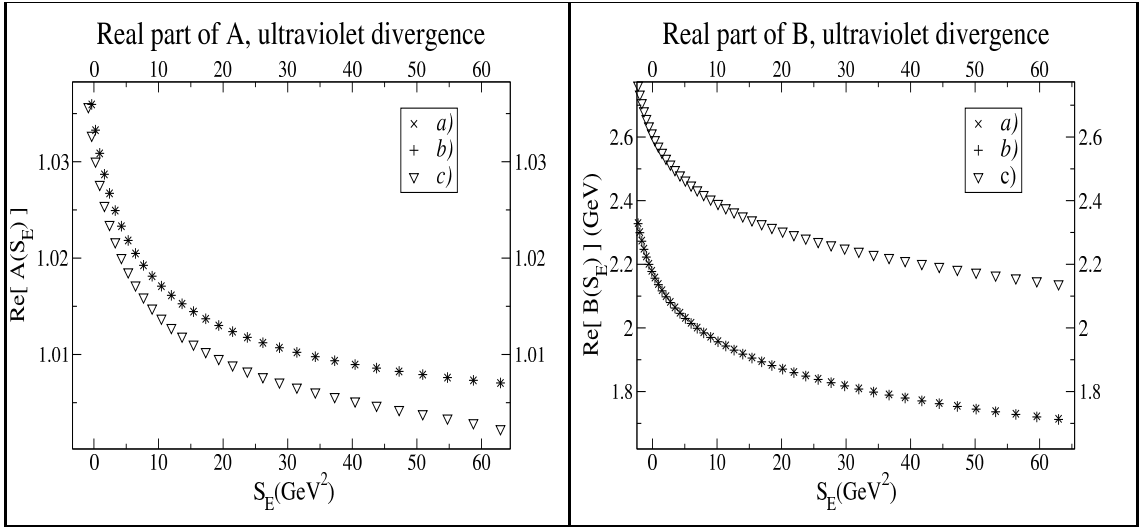}
\caption{Comparison of the numerical and analytical calculations of 
the MS-renormalized quark propagator after one iteration in an 
ultraviolet divergent case. $a)$ analytical subtraction in 
numerical integrations $b)$ with tail contribution 
and MS subtraction, $c)$ without tail contribution or 
MS subtraction. \label{sdlcx}}
\end{figure}

  The model gluon propagator is 
\beq
\label{gluongeneric}
 D_{\mu \nu}(k) = \frac{-1}{k^2+i\epsilon}
\left( g_{\mu \nu}- \chi \frac{ k_{\mu}k_{\nu} }{k^2+i\epsilon}\right)D(k^2).
\eeq
The parameter $\chi = 0$ for a Feynman-like gauge see Eq.(\ref{feynmanlike}) 
 and $\chi=1$ for Landau gauge.
It has been shown\cite{bg} that renormalization group arguments 
yield an approximate relation between the renormalized coupling constant, 
the renormalized gluon propagator and the effective coupling:
\beq 
\label{alphaD}
 g^2_R D_R(k^2) & \approx &  4 \pi \alpha_{eff}(k^2),
\eeq
where the subscript $R$ denotes renormalized quantities, 
and $g_{eff}$ is the effective running coupling constant. One model
used the polynomial form
 \beq
\label{alpharunning}
\alpha_{eff}(k^2)= \sum_{l=1}^{N} (-1)^{c_l} \lambda_l 
\left( \frac{ s_0}{k^2+i\epsilon} \right)^{c_l},
\eeq
and the parameters of the model were chosen to fit the PDG\cite{pdg} 
values, shown in Fig.~\ref{alphafit}.
A sample result of the calculation using the iterative procedure is 
shown in Fig.~\ref{sdlcx} 

\begin{figure}
\begin{center}
\epsfig{file=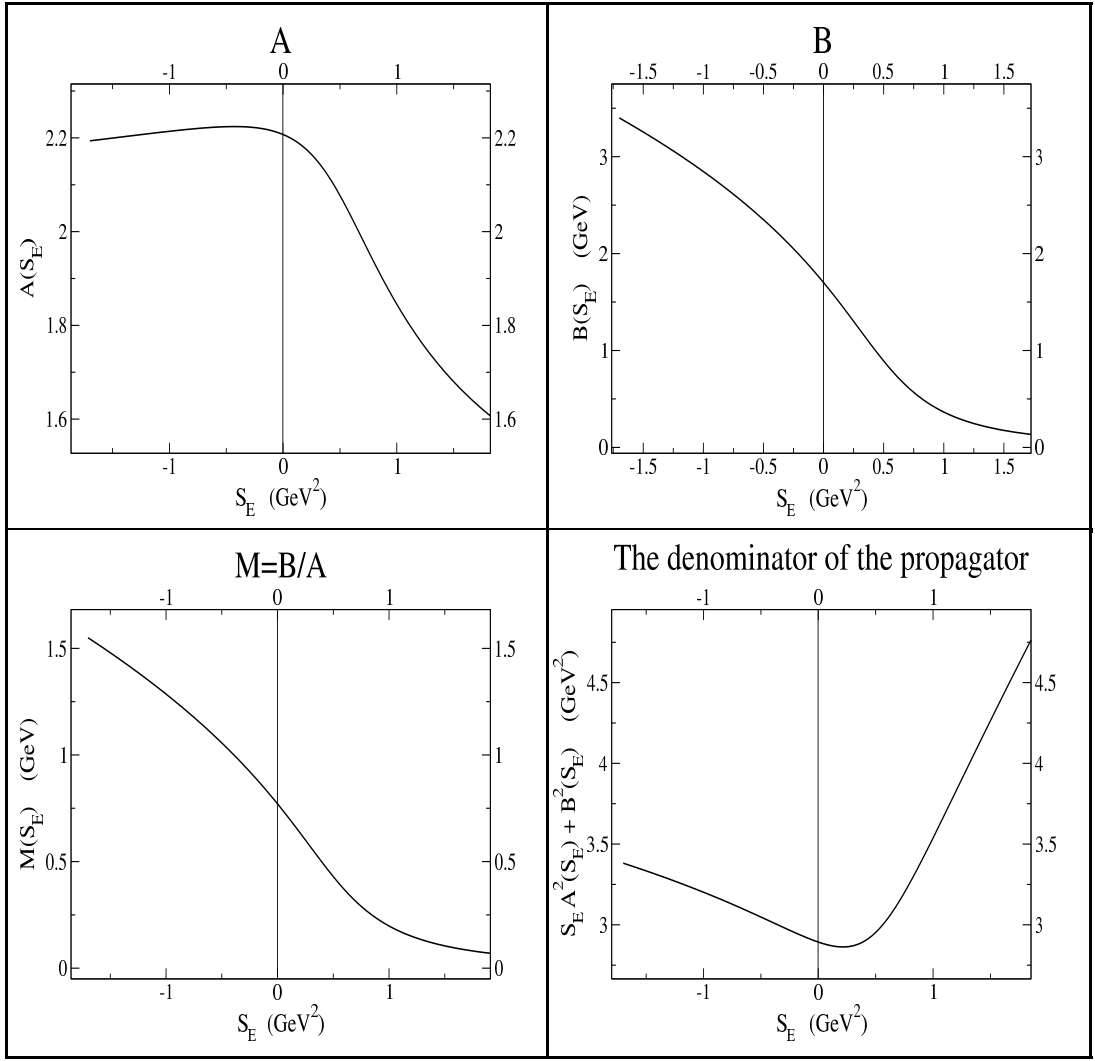,width=412pt,height=400pt}
\caption{Solutions to SDE in polynomial model. The left part of
the graphs ($S_E<0$) represents the time-like region.}
\label{sdesolution}
\end{center}
\end{figure}

In Fig.~\ref{sdesolution} the results of a calculation of the light cone
SDE are shown, with $S_E \equiv Q^2$, the Euclidean momentum squared.
The parameters for this calculation gave a reasonably good
fit to $f_\pi$ and $<:\bar{q}q:>$, but a rather large value for the mixed
condensate. It has a strong infra-red enhancement.
The results for the effective quark mass are too large at low momentum,
but the evolution in $Q^2$ is reasonable. A most interesting aspect of
this calculation is the ability to calculate the quark propagator in the
time-like region.

The polynomial model used to obtain the results in Figs.~\ref{alphafit},
\ref{sdesolution} describes an infrared-enhanced gluon propagator. 
Although this can give confinement,
it cannot give the QCD nonperturbative midrange attraction. On the other
hand the instanton model can give the strong midrange attraction, even
though it does not give the observed confinement. The instanton form
of the gauge color field is obtained in a SU(2) Euclidean classical
model\cite{bel75}
\begin{eqnarray}
\label{instanton}
   A^{inst}_{\mu(x) a} & = & \frac{2 \eta_{a\mu\nu} x_\nu}{x^2 + \rho^2} 
 \\ \nonumber
     G^{inst}(x) \cdot G^{inst}(x) & = & \frac{192 \rho^4}{(x^2 + \rho^2)^4}
\end{eqnarray}
where $\eta_{a\mu\nu}$ are given in Ref.\cite{th76} and for $\rho$, the 
instanton size, the instanton liquid model\cite{ss} is used. The quark
propagator in the instanton/anti-instanton medium, $S_I$, has been derived 
in this model\cite{pob89}, giving
\beq
\label{pob}
     S_I(p) & = & \left(\not\!pA_I(p^2)-B_I(p^2)\right)^{-1} \nonumber \\ 
            A_I(P) & = & 1 \nonumber \\
            B_I(P) & = & K P^2 f^2(\rho P/2)  \\
            f(z) & = & \frac{2}{z} -(3 I_0(z) + I_2(z)) K_1(z), \nonumber
\eeq
where $K \approx 0.29 \mbox{ GeV}^{-1}$, the instanton density $\rho 
\approx 1.667 \mbox{ GeV}^{-1}$, $P = \sqrt{-p^2}$, 
and the $I's \mbox{ and } K_1$ are 
modified Bessel functions of the first and second kind, respectively. 

  In Ref.~\cite{kl02} a number of calculations were done with both the
polynomial infra-red enhancement and the instanton contribution, with
various choices of the form to check for double-counting. See that
references for the results and details on how the calculation is done.

  This approach is most promising when coupled with the methods for
using SDE solutions to get mesonic properties as in 
Refs.~\cite{mr97,bur97,mt99,mt00}. In the previous work, in which
the key idea given by Eq.(\ref{SDBS}) is used, the use of instant form
SDE solutions limits the work to low momentum transfer, while with the
light cone solutions described in this section all momentum transfer 
can be studied. This is an exciting prospect for future research in
light cone physics.

\section{Deeply Virtual Compton Scattering and Skewed Parton Distributions}

   In recent years there has been a great interest in deeply virtual
compton scattering (DVCS) arising from the use of this reaction for
studying generalized parton distributions\cite{ji97,rad97}, also called
off-forward or skewed parton distributions (SPDs). See Ref.\cite{md03} for
a recent review with references. Virtual Compton
scattering from a hadron, i.e., the e + e' $\rightarrow \gamma$ reaction 
on the hadron, is the reaction
\beq
\label{vcs}
               \gamma^*(q) + H(P) &=& \gamma(q') + H(P'),
\eeq
where q (q') are the momenta of the incoing virtual (outgoing real)
photons and P (P'=P-$\Delta$) are the momenta of the initial (final) 
hadron, H. The DVCS amplitude is
\beq
\label{vcsamp}
   T^{\mu\nu} &=& i\int d^4x e^{-iq\cdot x}<H(P')|T[J^\mu_{em}
 J^\nu_{em}]|H(P)>,
\eeq
with T the time-ordered product of the electromagnetic currents.
 
\begin{figure}
\begin{center}
\epsfig{file=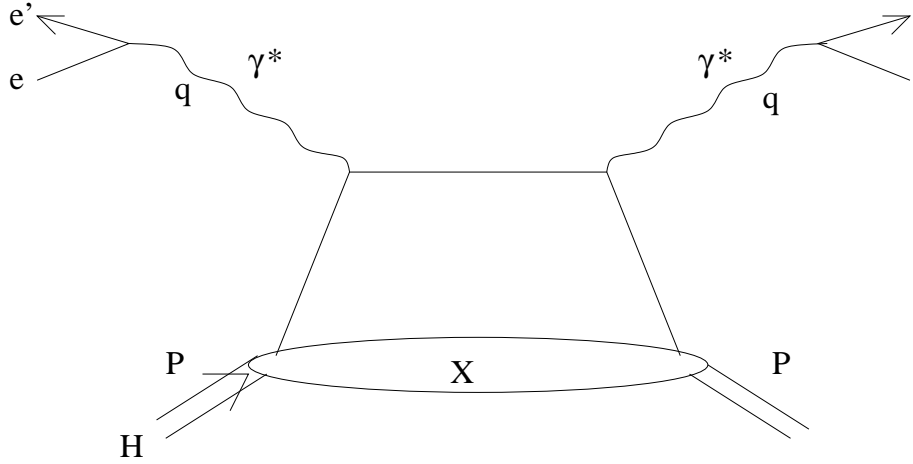,width=12 cm}
\caption{Handbag diagram for DIS cross section} 
\label{dis}
\end{center}
\end{figure}

  DVCS is the high $Q^2=-q^2$ limit of VCS, and has some of the physical
as well as mathematical aspects of deep inelastic scalleting (DIS). A 
striking aspect of inelastic scattering with a hadronic target H
\beq
\label{inelastic}
              e + H & \longrightarrow & e' + hadrons
\eeq
is that the differential cross section factorizes
\beq
\label{insigma}
        \frac{d^2 \sigma^{DIS}}{dE'd\Omega} &=&\frac{\alpha^2}{q^4}{E'}{E}
   L_e^{\mu\nu}W_{\mu\nu},
\eeq
where $L_e^{\mu\nu}$ is the leptonic tensor
\beq
\label{lepton}
       L_e^{\mu\nu} &=& \frac{1}{2}Tr[(\not\!k' + m_e)\gamma^\mu
 (\not\!k + m_e)\gamma^\nu],
\eeq
with $(k^{\mu}, k^{'\mu})$ the (incoming, outgoing) electron 
four-momenta, $\alpha^2=e^2/4\pi$and $W_{\mu\nu}$ is the hadronic tensor.  
For DIS the ``handbag'' diagram shown in Fig.~\ref{dis}
dominates. In the diagram $H$ is the hadron target and the sum is taken 
over intermediate states $X$, expressing the hadronic tensor as
\beq
\label{hadrontensor}
      W^{\mu\nu}(P,q) &=& \frac{1}{2\pi M}\int d^4x e^{iq\cdot x}
 <H(P,\lambda)|[J^\mu_{em}(x),J^\nu_{em}(0)]|H(P,\lambda)>
\eeq
in terms of the matrix element of the current commutator n the hadronic
state with spin index $\lambda$. For more than two decades DIS has been
used to learn about parton distributions. (See, e.g., Ref\cite{close79})

\begin{figure}
\begin{center}
\epsfig{file=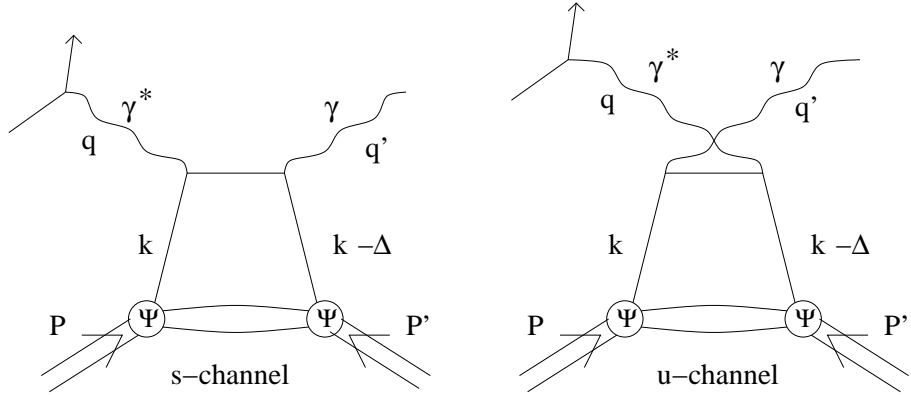,width=12 cm}
\caption{Handbag diagrams for DVCS amplitude. a) s-chanel; b) u-channel.} 
\label{handbag}
\end{center}
\end{figure}

   For DVCS the amplitude (not the cross section as in DIS) is dominated
by the handbag diagrams, shown in Fig.~\ref{handbag}.
In the diagram the target is shown as $\Psi$, the B-S amplitude of the
hadron, which gives the vertex of H, a quark, and the residual system.
After a review of the kinematics we give a brief review of DVCS on a
nucleon and review recent developements on skewed parton distributions
of the pion, where newly developed B-S amplitudes can be used to
compare with experiment.

\subsection{Kinematics for DVCS. Skewness}

   With the notation for  a light cone four vector $V^\mu = 
(V^+,V^-,\vec{V}_\perp)$, choosing the frame with $\vec{P}_\perp = 0$, 
and defining the four momentum transfer $\Delta=P-P'$,
the kinematics of VCS with a hadron target of mass $M$ is 
\beq
\label{PP}
P &=&\biggl[P^+, \frac{M^2}{P^+}, 0_\perp \biggr],\nonumber\\
P'&=&\biggl[(1-\xi)P^+,
\frac{M^2+\Delta^{2}_{\perp}}{(1-\xi)P^+}, -\Delta_\perp\biggr],
\eeq
\beq
\label{del}
\Delta &=& P - P' =\biggl[
\xi P^+, \frac{\Delta^2+\Delta^2_\perp}{\xi P^+}, \Delta_\perp \biggr],
\eeq
with  
\beq
\label{skewness}
  \xi & \equiv & \Delta^+/P^+
\eeq 
the skewedness parameter describing the asymmetry in plus momentum.
The squared momentum transfer is
\beq
\label{del2}
 t &=& \Delta^2 = 2 P\cdot\Delta=-\frac{\xi^2 M^2 + \Delta^2_\perp}{1-\xi}.
\eeq
 
  In a frame where the incident spacelike photon has $q^+=0$
\beq
\label{qq}
q &=& \biggl[0, \frac{({\bf q}_\perp + \Delta_\perp)^2}{\xi P^+}
+ \frac{\xi M^2 +\Delta^2_\perp}{(1-\xi) P^+},
{\bf q}_\perp \biggr],\nonumber\\ 
q' &=& \biggl[\xi P^+, 
\frac{({\bf q}_\perp + \Delta_\perp)^2}{\xi P^+},
{\bf q}_\perp +\Delta_\perp\biggr]. 
\eeq
In deeply virtual Compton scattering (DVCS) where $Q^2=-q^2$ is large 
compared to the mass $M$ and $-t$, one obtains
\beq
\label{Bj}
\frac{Q^2}{2P\cdot q} &\simeq& \xi,
\eeq
i.e. $\xi$ plays the role of the Bjorken variable in DVCS. For a fixed
value of $-t$, the allowed range of $\xi$ is given by
\beq
\label{range}
 0\leq\xi\leq\frac{(-t)}{2M^2}\biggl(
\sqrt{1 + \frac{4M^2}{(-t)}}-1\biggr).
\eeq

\subsection{DVCS on the Nucleon}

   There has been a great deal of interest in elastic and inelastic form
factors of the nucleon for many years for tests of quark models and
quark distributions. The interest in DVCS was the realization that new
information on quark structure of the nucleon can be obtained, and it
can be directly related to the elastic structure functions.
First recall the elastic form factors of the nucleon, $F_1,F_2$,
\beq
\label{nff}
  <n(P')|J^{em}_\mu|n(P)> &=& \bar{N}(P')[\gamma_\mu F_1(q^2)
 +i\sigma_{\mu\nu}\frac{q^\nu}{2M}F_2(q^2)]N(P),
\eeq
with $N(P)$ the nucleon spinor, the axial form factors
\beq
\label{axialff}
<p(P')|A^1_\mu+iA^2_\mu|n(P)> &=& \bar{N_p}(P')[\gamma_\mu \gamma_5 G_A(q^2)
 +q_\mu \gamma_5 H_A(q^2)]N_n(P),
\eeq
where $A^i_\mu = \bar{N}\gamma_\mu \gamma_5 \frac{\tau^i}{2}$ is the axial 
vector current, and the pseudoscalar form factor
\beq
\label{psff}
     <n(P')|\frac{\gamma_5}{2M}|n(P)> &=& \bar{N}(P')G_{P}(q^2)N(P).
\eeq

   The VCS amplitude with the handbag diagrams, Fig.~\ref{handbag}, have
the form 
\beq
\label{vscnuc}
  T^{\mu\nu} &=& i \int \frac{d^4 k}{(2\pi)^4} Tr[\gamma^nu S(k+q) \gamma^mu
+\gamma^mu S(k-q) \gamma^nu]\mathcal{F}(k),
\eeq
where $S$ is the quark propagator and 
\beq
\label{quarkdist}
   \mathcal{F}(k) &=& \int e^{ikx} d^4x <N(p')|\bar{\psi}(0)\psi(x)|N(P)>.
\eeq
When the DVCS amplitude is expressed in light cone variables, the quark
matrix element $\mathcal{F}$ is expressed in terms of skewed quark 
distributions which in the notation of Ref.\cite{ji97} are to twist two
\beq
\label{jiskew}
   \int \frac{dz_-}{2\pi}e^{iz_- x} <N(p')|\bar{\psi}(0)\gamma^\mu
 \psi(z_-)|N(P)> &\simeq & H(x,\Delta^2,\xi) \bar{N}(P')\gamma^\mu N(P)
\nonumber \\ 
  &&+ E(x,\Delta^2,\xi) \bar{N}(P')\frac{i\sigma^{\mu\nu}\Delta_\nu}
{2M} N(P) \nonumber \\
    \int \frac{dz_-}{2\pi}e^{iz_- x} <N(p')|\bar{\psi}(0)\gamma^\mu
\gamma_5 \psi(z_-)|N(P)> &\simeq & \tilde{H}(x,\Delta^2,\xi) 
\bar{N}(P')\gamma^\mu \gamma_5 N(P) \nonumber \\ 
  && +\tilde{E}(x,\Delta^2,\xi) \bar{N}(P')\frac{\gamma_5\Delta^\nu} N(P)
\eeq
These skewed distributions satisfy the sum rules
\beq
\label{skewsum}
  \int_{-1}^{1} dx H(x,\Delta^2,\xi) &=& F_1(\Delta^2) \nonumber \\
  \int_{-1}^{1} dx E(x,\Delta^2,\xi) &=& F_2(\Delta^2) \nonumber \\
  \int_{-1}^{1} dx \tilde{H}(x,\Delta^2,\xi) &=& G_A(\Delta^2) \\
  \int_{-1}^{1} dx \tilde{E}(x,\Delta^2,\xi) &=& G_P(\Delta^2). \nonumber
 \eeq
 
  In recent years there have been studies of higher twist effects for DVCS.
For a recent review of DVCS with a nucleon target and references to recent
work see Ref.~\cite{bmk02}, and a light cone wave function representation
of DVCS is discussed in Ref.~\cite{bdh01}.
 There is a great interest in new experiments
to measure skewed quark distributions. These were discussed at the 
international workshop Light cone 2002\cite{lc02}.

   We now turn to DVCS and skewed quark distribution on the pion.

\subsection{DVCS on the Pion}

   The skewed quark distributions of the pion,${\cal F}_{\pi}(\xi,x,t)$,
 are defined by the integral
\beq
\label{SPD}
\int\frac{dz^-}{4\pi}e^{i x P^+z^-/2}
\langle P'|{\bar\psi}(0)\gamma^+\psi(z)|P\rangle|_{z^+={\bf z}_{\perp}=0} 
&=&{\cal F}_{\pi}(\xi,x,t)(P+P')^+,
\eeq     
where  $z=(z^+,z^-,{\bf z}_\perp)$ in a light front representation. 
Note that the path-ordered exponential of the gauge field,
${\cal P}\exp[i\int z^{\mu}A_{\mu}]$, required by gauge invariance\cite{rad97} 
in Eq.~(\ref{SPD}) does not appear in the light front gauge $A^+=0$.
Recalling the definition of the pion form factor, $F_{\pi}(t)$,
\beq
\label{EM}
J^+(0)\equiv\langle P'|{\bar\psi(0)}\gamma^+\psi(0)|P \rangle
&=&F_{\pi}(t)(P+P')^+,
\eeq
as one can see from Eqs.~(\ref{EM}) and~(\ref{SPD}), 
the ${\cal F}_{\pi}$ involves one less integration than
the form factor $F_{\pi}$ due to nonlocality of the current matrix 
element. The SQDs display characteristics of
the ordinary(forward) quark distribution in the limit of $\xi\to 0$
and $t\to 0$, on the other hand,  the first moment of the SQDs
is related to the form factor by the following sum 
rules~\cite{ji97,rad97}:
\beq
\label{sum}
\int^1_{0}dx\; {\cal F}_{\pi}(\xi, x, t) &=& F_\pi(t),
\eeq
where ${\cal F}_{\pi}(\xi, x, t)=e_u{\cal F}^u_{\pi}(\xi, x, t)
-e_d{\cal F}^{\bar d}_{\pi}(\xi, x, t)$ and we assume 
isospin symmetry($m_u=m_{\bar d}$) so that  
${\cal F}^u_{\pi}(\xi, x, t)={\cal F}^{\bar d}_{\pi}(\xi, x, t)$.
Note that Eq.~(\ref{sum}) is independent of $\xi$, which provides
important constraints on any model calculation of the SQDs.

\begin{figure}[h]
\begin{center}
\epsfig{file=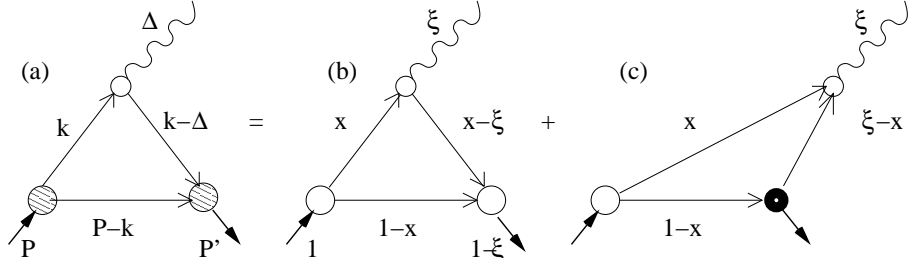,width=12 cm}
\caption{ Diagrams for SQDs in different kinematic regions for the
case $\xi>0$: The covariant diagram (a) corresponds to the sum of
the LF valence diagram (b) defined in $\xi< x<1$ region 
and the nonvalence diagram (c) defined in $0< x<\xi$ region.
The large white and black blobs at the meson-quark vertices in (b) and
(c) represent the ordinary LF wave function and the nonvalence 
wave function vertices, respectively.  
The small white blob at the quark-gauge boson 
vertex indicates the nonlocality of the vertex.} 
\label{highFock}
\end{center}
\end{figure}

The diagrams for the SQDs are shown in Fig.~\ref{highFock}. In the
range of x, $0<x<1$, for $x>\xi$ the BS amplitude is the standard one
that determines the pion form fator, as shown in Fig.~\ref{highFock}(b),
while for $x<\xi$ an analytic continuation, called the nonvalence BS
amplitude is needed, as shown in Fig.~\ref{highFock}(c).
 
The B-S amplitudes (see Sec. 4) needed for the calculation of the SQDs are
solutions to 
\beq\label{SDtype1}
(M_{\xi}^2-{\cal M}^2_0)\chi(x_i,{\bf k}_{i\perp})
&=&\int [dy][d^2{\bf l}_\perp]
{\cal K}(x_i,{\bf k}_{i\perp}; y_j,{\bf l}_{j\perp})
\chi(y_j,{\bf l}_{j\perp}),
\eeq
where ${\cal K}$ is the B-S kernel, $M_{\xi}^2=M^2/(1-\xi)$,${\cal M}^2_0=
(m^2+{\bf k}^2_{\perp})/(1-x) - (m^2 + {\bf k}^2_{\perp})/(\xi-x)$,
and $\chi(x_i,{\bf k}_{i\perp})$ is the B-S amplitude. Defining the solution
to Eq.(\ref{SDtype1}) when $x>\xi$, which is the familiar B-S amplitude, as
the valence amplitude, the valence and nonvalence B-S solutions are
\beq
\label{BSamp1}
        \chi^{val} &=& \chi \; ({\rm x \; > \xi}) \nonumber \\
        \chi^{nv} &=& \chi \; ({\rm x \; < \xi}) \\ 
\eeq
In the light front quark model\cite{JC01} the B-S amplitudes in the two
regions are related by
\beq\label{SDnonval}
(M_{\xi}^2-{\cal M}^2_0)\chi^{nv}(x_i,{\bf k}_{i\perp})
&=&\int [dy][d^2{\bf l}_\perp]
{\cal K}(x_i,{\bf k}_{i\perp}; y_j,{\bf l}_{j\perp})
\chi^{val}(y_j,{\bf l}_{j\perp}) \; .
\eeq
The contribution from the process shown in Fig.\ref{highFock}(b) is given by
the ${\rm x \; > \xi}$ B-S amplitude, with the valence part of 
${\cal F}_{\pi}(\xi,x,t)$ given by
\beq\label{jv}
{\cal F}^{val}_\pi(\xi,x,t)&=&\frac{N_c}{(P+P')^+}\int^1_\xi
\frac{dx}{16\pi^3}\frac{\delta(x-k^+/P^+)4P^+}{x(1-x)x'(1-x')} \nonumber \\
 &&  \int d^2{\bf k}_\perp
\chi^{val}(x,{\bf k}_\perp) ({\bf k}_\perp\cdot {\bf k}'_\perp +m^2)
\chi^{' val}(x',{\bf k}'_{\perp}),
\eeq   
with the internal momenta of the (struck) quark for the final state 
given by
\beq\label{xpkp}
x'&=& \frac{x-\xi}{1-\xi},\;\;
{\bf k'}_\perp={\bf k}_\perp + \frac{1-x}{1-\xi}\Delta_\perp.
\eeq
The contribution from the nonvalence part of  
${\cal F}_{\pi}(\xi,x,t)$, the process shown in Fig.\ref{highFock}(c),
is given by the ${\rm x \; > \xi}$ B-S amplitude, with the form
\beq\label{jnv}
{\cal F}^{nv}_\pi(\xi,x,t)&=&\frac{N_c}{(P+P')^+}\int^\xi_0
\frac{dx}{16\pi^3}\frac{\delta(x-k^+/P^+)4P^+}{x(1-x)x'(1-x')} \nonumber \\
 && \int d^2{\bf k}_\perp \chi^g\chi^{val}(x,\vec{k}^{''}_\perp) 
H(x,{\bf k}_\perp) \chi^{nv}(x',{\bf k}'_{\perp}) \; ,
\eeq
with $H(x,{\bf k}_\perp)$ a function of kinematic variables given in
Ref\cite{cjk01a}, $\chi^g$ is the quark-gauge boson vertex shown as
the small white blob in the upper vertices of Fig~\ref{highFock},
and $\vec{k}^{''}_\perp =\vec{k}_\perp+x^{''}\Delta_\perp$
and $\chi^{nv}$

The contribution from this nonvalence region ${\rm x \; < \xi}$ 
is obtained by making approximate use of the constraint
\beq\label{2sum}
F_\pi (t)&=&\int^1_\xi dx\; {\cal F}^{val}_\pi(\xi, x,t)
+ \int^\xi_0 dx\; {\cal F}^{nv}_\pi(\xi, x,t),
\eeq
so the B-S amplitude $\chi^{val}(x,{\bf k}_\perp)$ is not used. The
valence B-S amplitude is modeled by a light cone harmonic oscillator:
\beq\label{LFvertex}
\chi^{val}(x,{\bf k}_\perp) &=&\biggl(\frac{2\pi^{3/2}}{N_cM_o\beta^3}
\biggr)^{1/2}\exp(-{\bf k}^2/2\beta^2) \; .
\eeq
Substituting Eq.(\ref{LFvertex}) in Eq.(\ref{jv}) one obtains the valence 
contribution to the SQDs.

\begin{figure}[ht]
\begin{center}
\epsfig{file=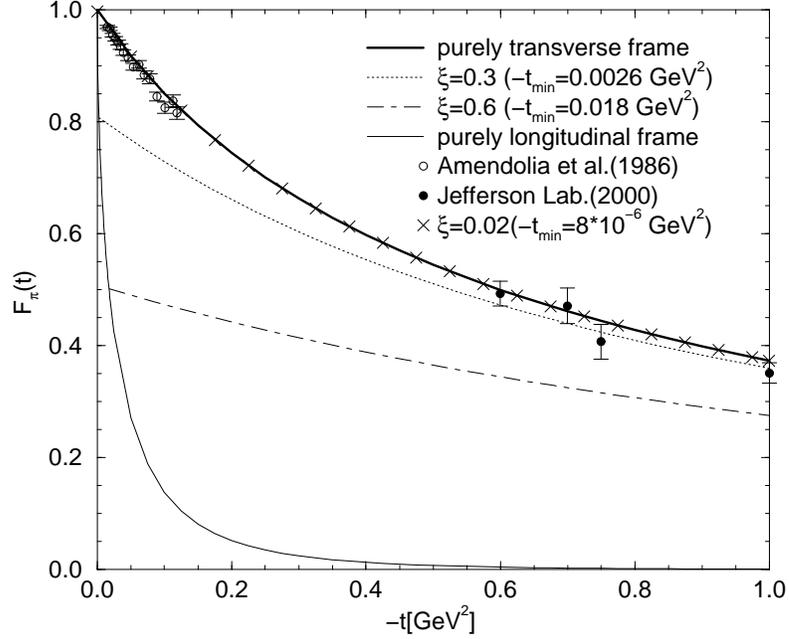,width=12 cm}
\caption{The valence contribution to the pion EM form factor with different
skewedness parameters $\xi$ compared with the experimental
data~\protect\cite{Amen,JLab}.\label{piform}}
\end{center}
\end{figure}

For $\xi = 0$ the pion form factor is given by the valence process. In
Fig.\ref{piform} the valence contribution to $F_\pi$ is shown. 
For values of $\xi > 0.3$ the nonvalence contribution is seen to become
significant.

\begin{figure}[p]
\centerline{\psfig{figure=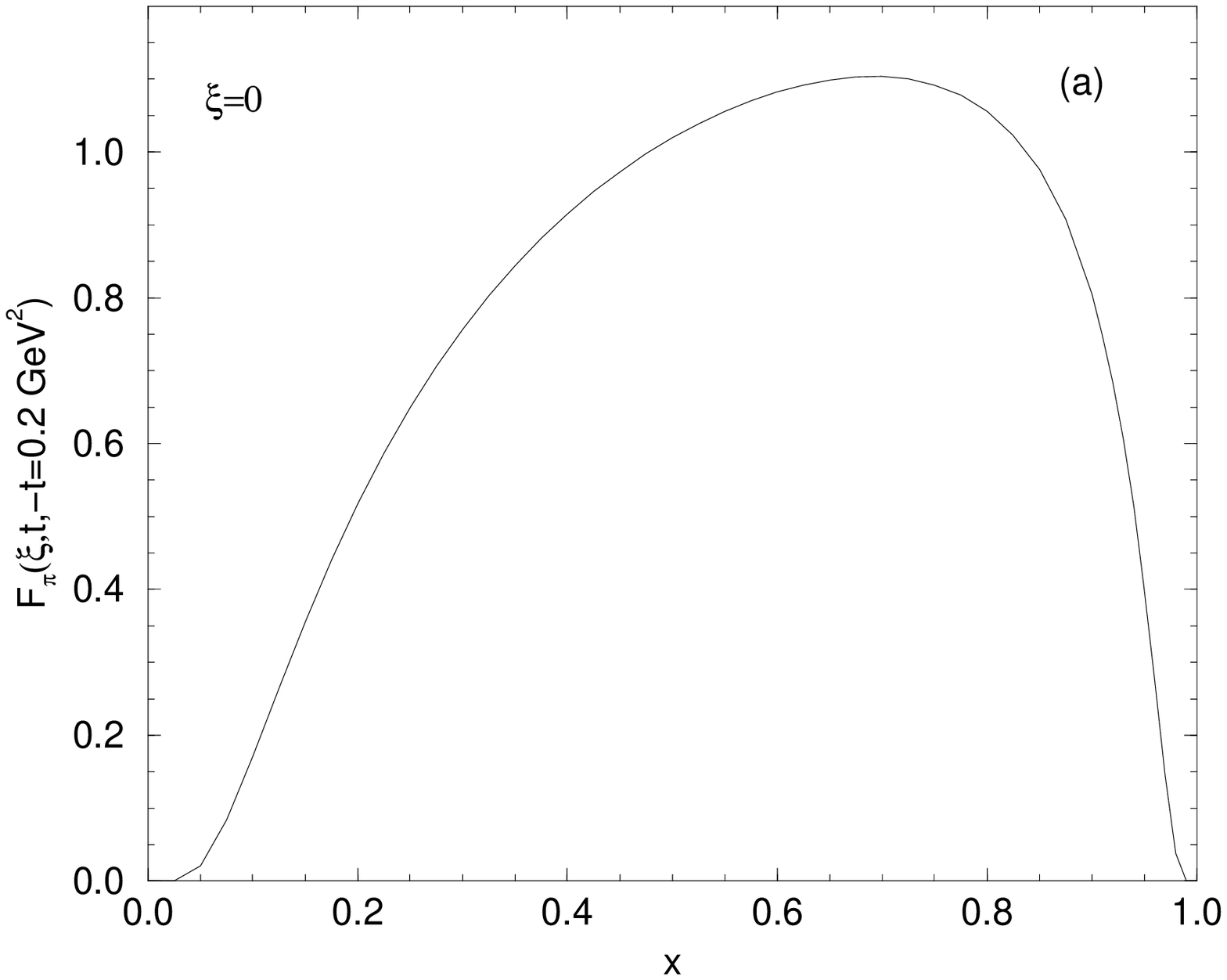,height=6.5cm,width=9cm}}
\centerline{\psfig{figure=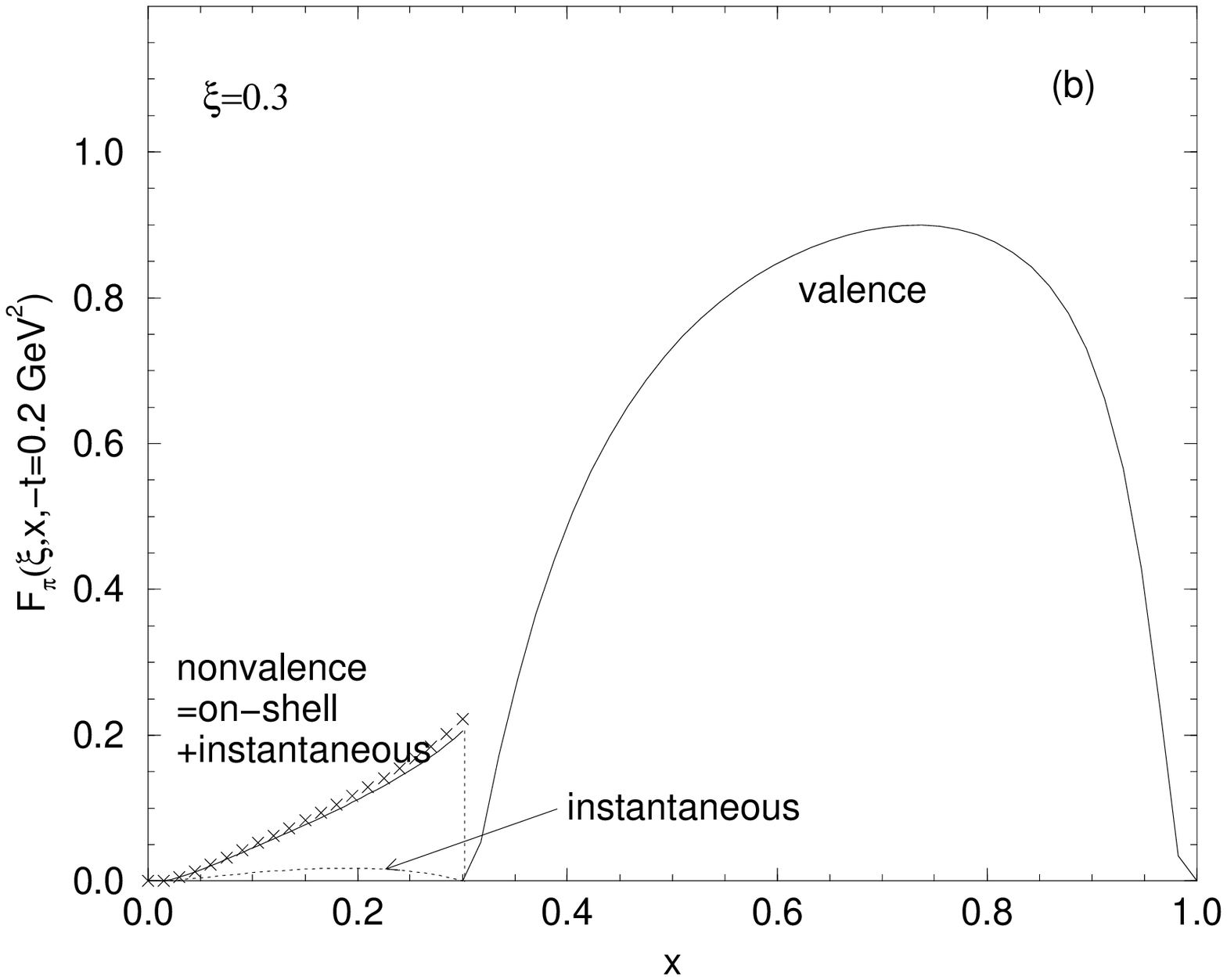,height=6.5cm,width=9cm}}
\centerline{\psfig{figure=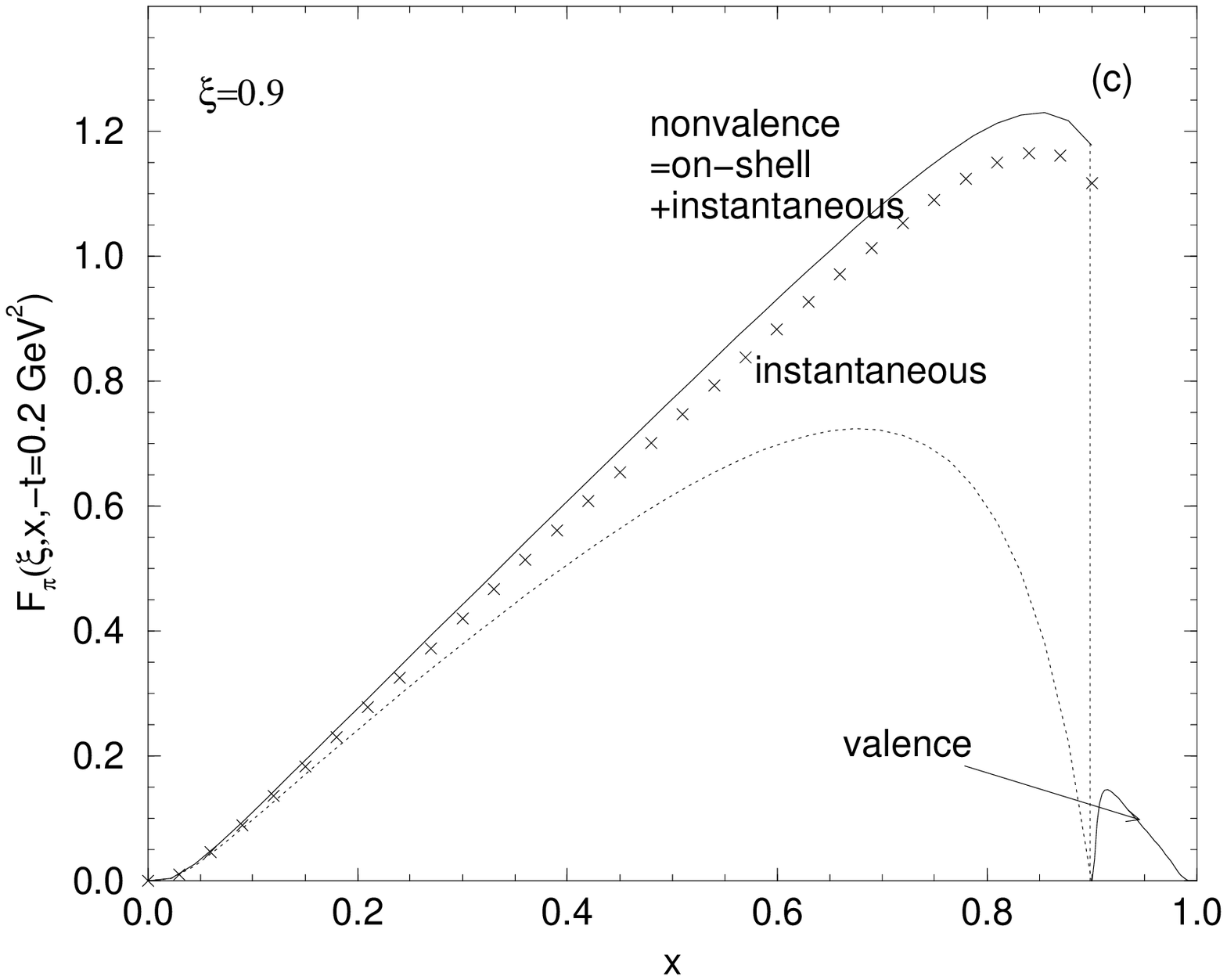,height=6.5cm,width=9cm}}
\caption{Skewed quark distributions of the pion at
$-t=0.2$ GeV$^2$ with $\xi=0$ in (a) ,0.3 in (b), and 0.9 in (c),
respectively. The solid [cross (x)] line in nonvalence contribution
represents the full result of using true [average] $G_\pi$ value and the
dotted line represents the instantaneous part of the nonvalence contribution.
\label{SQD02}}
\end{figure}

\begin{figure}[p]
\centerline{\psfig{figure=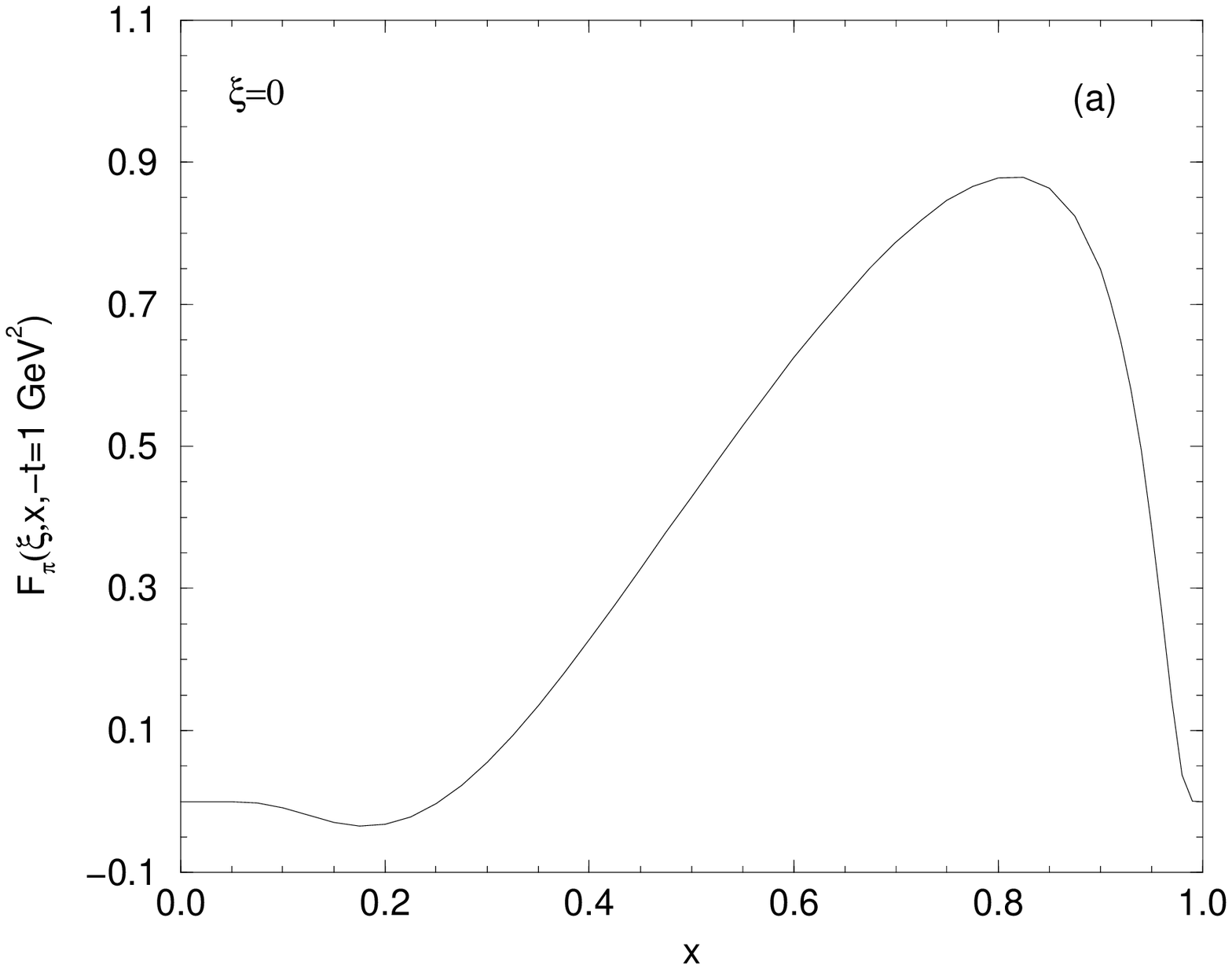,height=6.5cm,width=9cm}}
\centerline{\psfig{figure=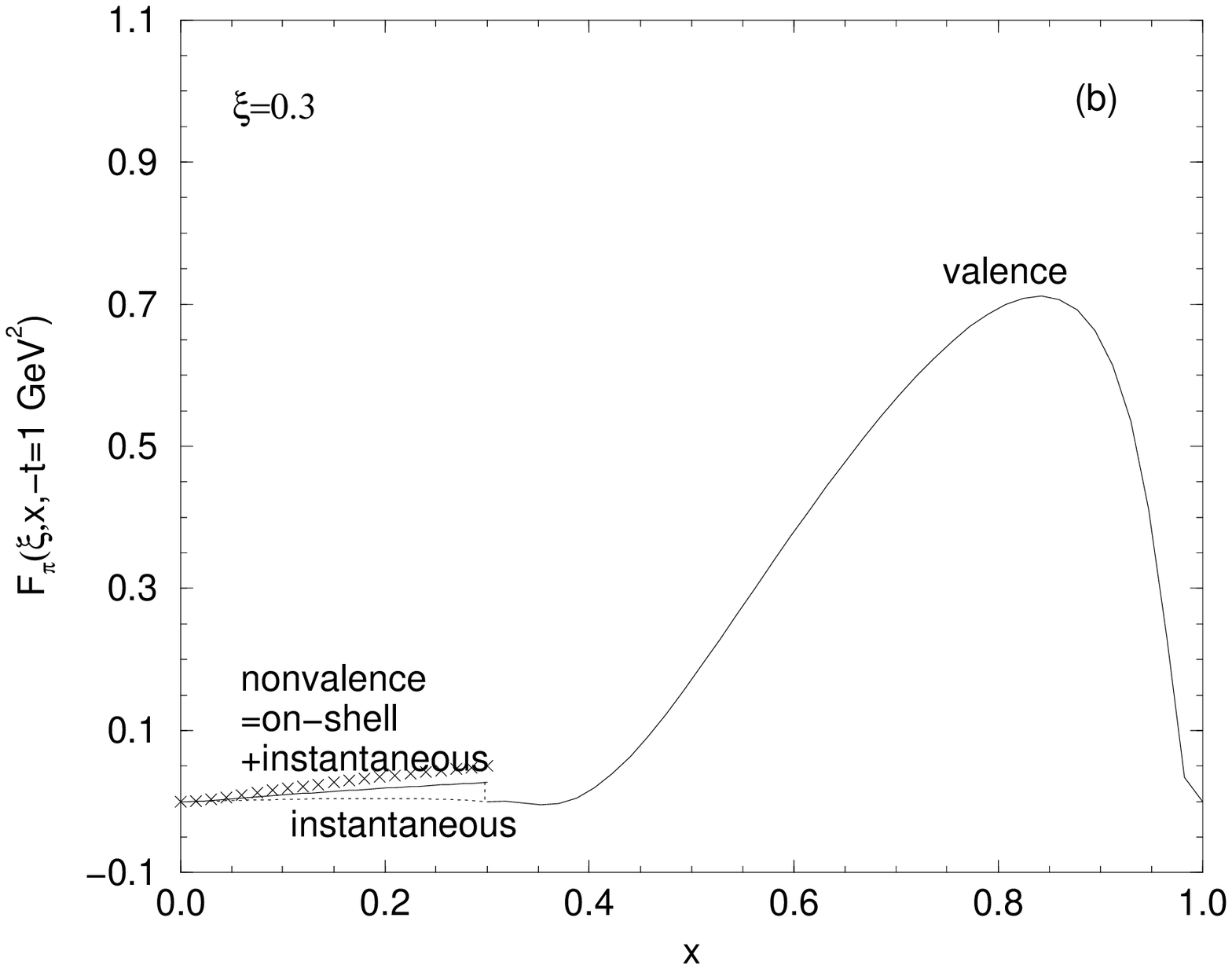,height=6.5cm,width=9cm}}
\centerline{\psfig{figure=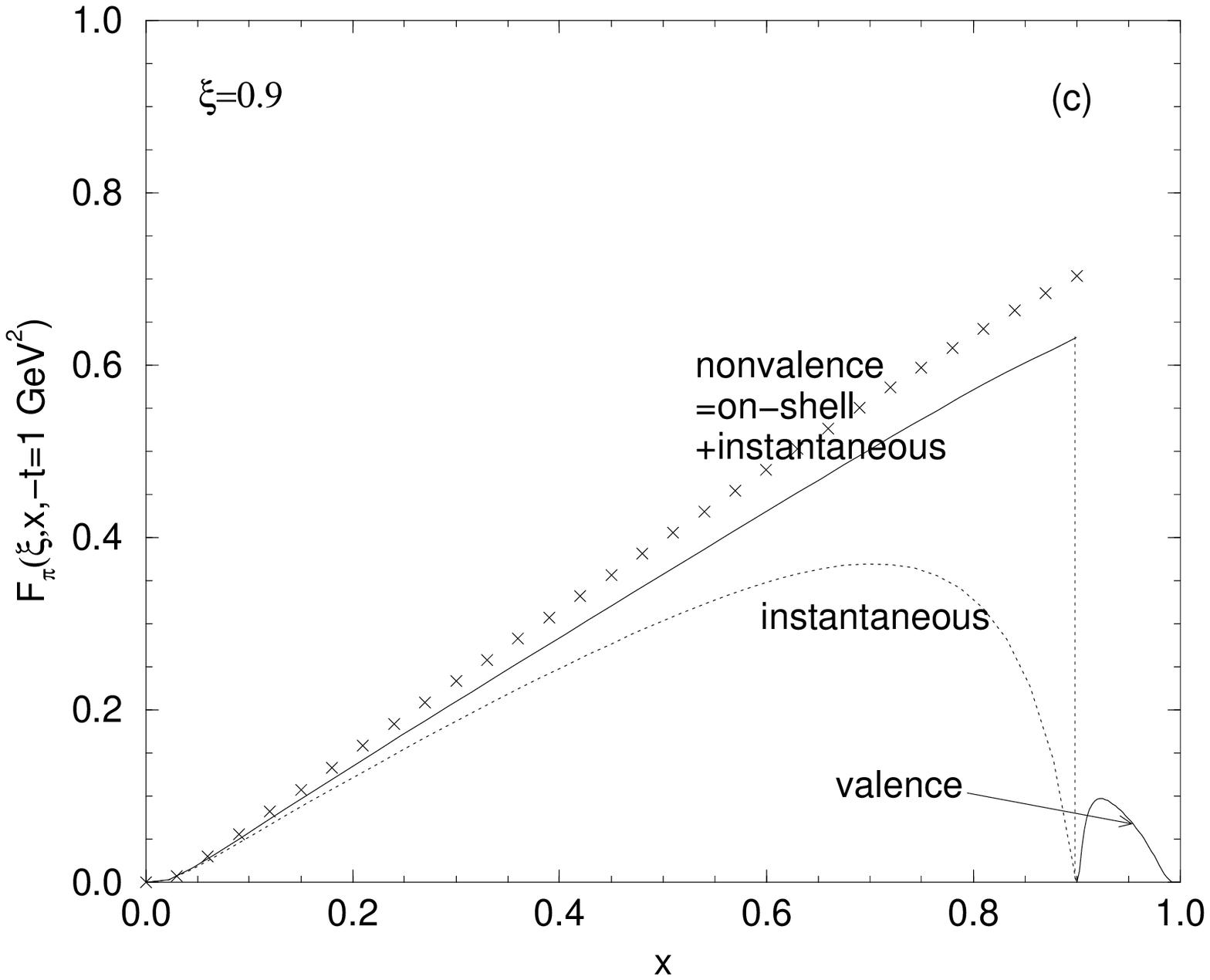,height=6.5cm,width=9cm}}
\caption{Skewed quark distributions of the pion at
$-t=1$ GeV$^2$ with
different $\xi=0$ in (a), 0.3 in (b), and 0.9 in (c), respectively.
The same notation is used as in Fig.~\protect\ref{SQD02}.
\label{SQD1}}
\end{figure}

In Figs.~\ref{SQD02} and~\ref{SQD1}, are shown the SQDs, 
${\cal F}_\pi(\xi,x,t)$, of the pion for fixed momentum transfer
$-t=0.2$ GeV$^2$ ($0\leq\xi\leq 0.92$) and
$-t=1.0$ GeV$^2$ ($0\leq\xi\leq 0.98$)
but with different skewedness parameters $\xi$, respectively.   
The solid and cross (x) lines in the nonvalence contributions 
are the exact solutions obtained from using the B-S equation to
obtain $\xi^{nv}$ for the model of Eq.(\ref{LFvertex}), and an
approximation based on Eq.(\ref{2sum}).
The SQDs at $\xi=0$ as shown in
Figs.~\ref{SQD02}(a) and~\ref{SQD1}(a) correspond to the ordinary 
quark distributions with vanishing nonvalence contributions.
The frame-independence of the model calculation is ensured by the
area under the solid lines(valence $+$ nonvalence) being equal to the
pion form factor at given $-t$.
As one can see from Figs.~\ref{SQD02}(b-c) and~\ref{SQD1}(b-c), while
the nonvalence contributions are small for small $\xi=0.3$, they
are large for large skewdness parameter $\xi=0.9$. 
It is interesting to note that the instantaneous
contribuitons (dotted lines in each figure) become more pronounced as 
$\xi\to\xi_{\rm max}$ for each $-t$.
While the instantaneous part of the nonvalence contribution vanishes
as $x \rightarrow \xi^-=lim_{\epsilon\rightarrow 0}(\xi-\epsilon)$ as 
shown in Figs.~\ref{SQD02} and~\ref{SQD1}, the net result
of ${\cal F}^{nv}_\pi(\xi,x,t)$ including the on-mass shell propagating
part does not
vanish as $x=\xi$ and consequently causes a discontinuity to the zero
value of ${\cal F}^{val}_\pi(\xi,\xi,t)$. 
However, such discontinuity at $x=\xi$ is just an artifact due to 
the difference in the $x\rightarrow\xi$ behavior between the gauge boson 
vertex ($\chi^g$ in Ref.\cite{cjk01a}) and the hadronic
vertex ($\chi^{val '}(x',{\bf k}'_\perp))$ in Eq.~(\ref{jv}) of 
our approximate model calculation. See the following subsection for
a discussion of $\chi^g$ and the continuity problem.
We have indeed confirmed that the discontinuity at $x=\xi$ does not
occur in the limit of a point hadron vertex as observed
in the QED calculation~\cite{bdh01}.
Thus, for a full analysis of DVCS satisfying factorization 
theorems, it would be necessary to solve the bound-state B-S 
equation similar to Eq.~(\ref{SDtype1}) for the gauge boson ($\chi^g$) as
well as for the hadron ($\chi^{val}$).

The main approximation in this calculation is the treatment of relevant 
operator ${\cal K}(x,{\bf k}_\perp;y,{\bf l}_\perp)$ in Eq.~(\ref{SDnonval}) 
connecting the one-body to three-body sector  by taking a constant  
$G_\pi$ for the quantity $G_\pi\equiv\int[dy][d^2{\bf l}_\perp]
{\cal K}(x,{\bf k}_\perp;y,{\bf l}_\perp)\chi^{val}(y,{\bf l}_\perp)$,
which in general depends on $x$ and ${\bf k}_\perp$.
The reliability of this approximation was checked by examining
the frame-independence of the numerical results and  
using the sum rule given by Eq.~(\ref{2sum}).
This method seems useful for the present
study of the relation between SQDs and the form factor in the 
nonperturbative regions.

\subsubsection{Continuity of Pion DVCS Generalized Parton Distributions}   

\begin{figure}[ht]
\centerline{\psfig{figure=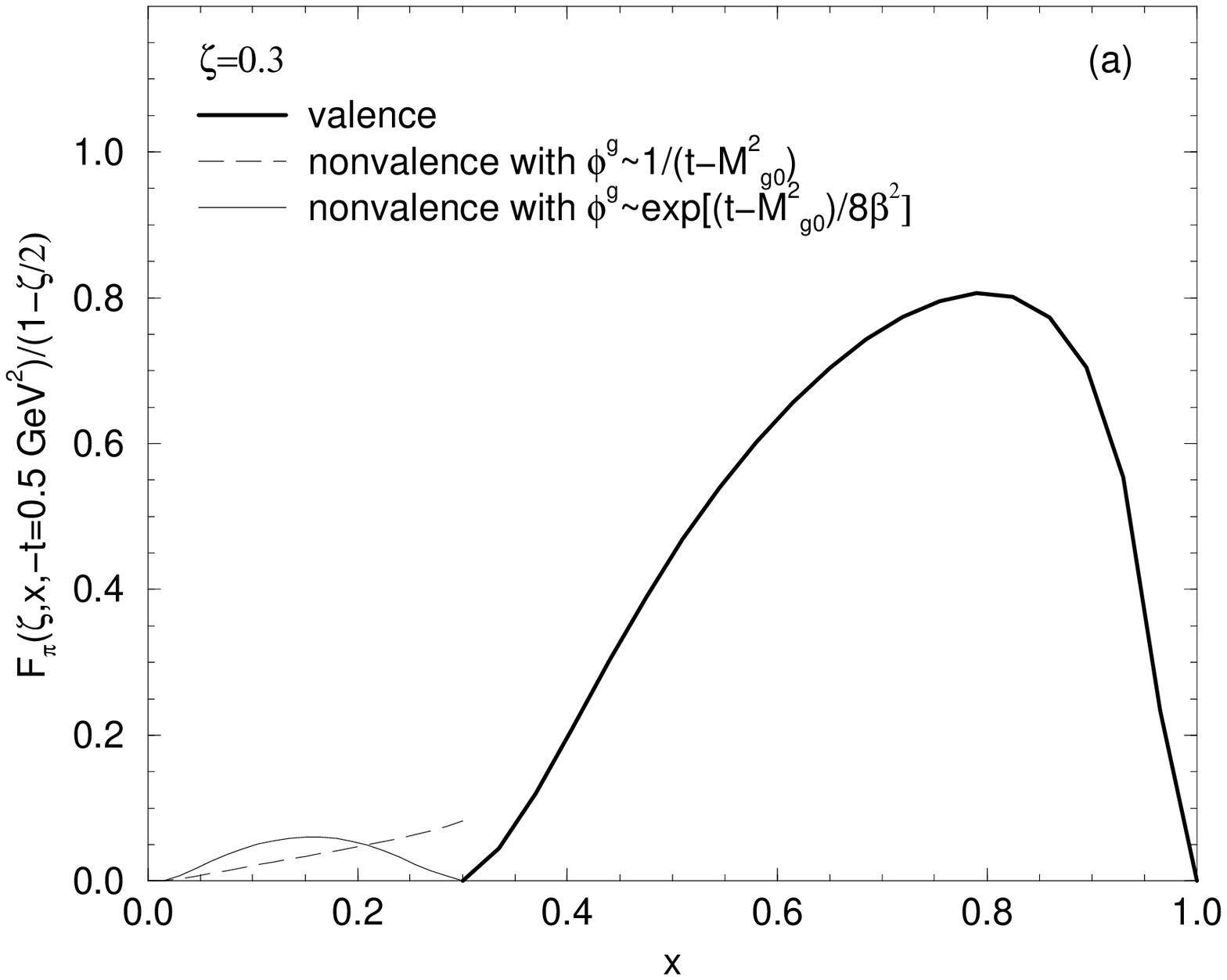,height=6.5cm,width=9cm}}
\centerline{\psfig{figure=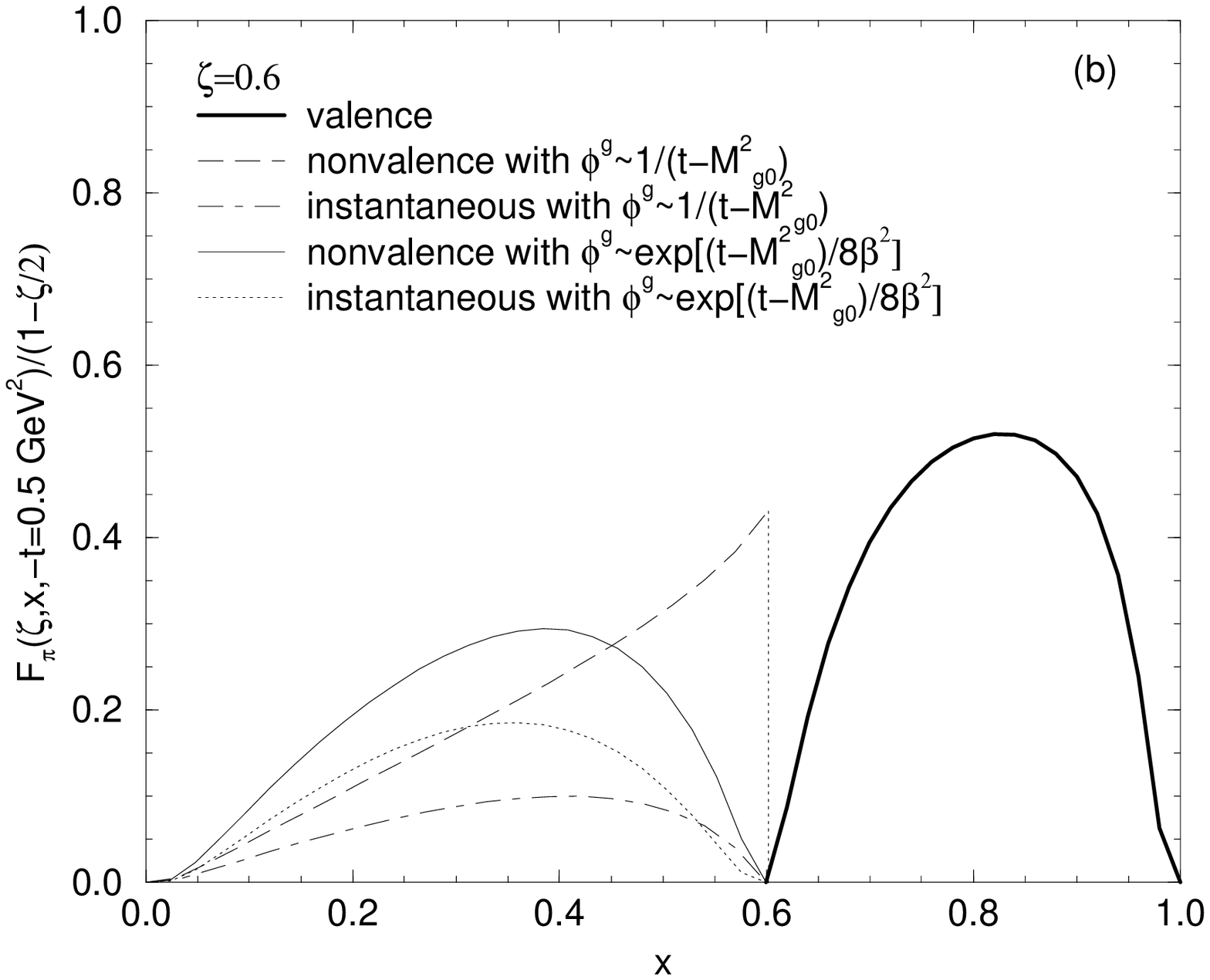,height=6.5cm,width=9cm}}
\caption{Generalized parton distributions of the pion with modified
gauge boson wave function at
$-t=0.5$ GeV$^2$ with $\xi=0.3 $ in (a) and 0.6 in (b).
\label{SDQCONT}}
\end{figure}

   The problem with continuity, discussed in the previous section, is
the treatment of the gauge boson vertex, $\chi^g$ in the nonvalence region, 
$\xi \geq x$; i.e., the white blob in Fig.\ref{highFock}(c). 
In Ref\cite{cjk01a} the form used was
\beq
\label{gaugewf1}
 \chi^g(x^{''},\vec{k}^{''}_\perp) &=& \frac{1}{\Delta^2 - M^{'' 2}} \;,
\eeq
with $x^{''} = x/\chi, \vec{k}^{''}_\perp =\vec{k}_\perp+x^{''}\Delta_\perp$
and $M^{'' 2} = (\vec{k}^{'' 2}_\perp+m^2)/x^{''} +
(\vec{k}^{'' 2}_\perp+m^2)/(1-x{''})$. 
In \cite{cjk02} an effective light cone wave function was used for the gauge
boson wave function:
\beq
\label{gaugewf2}
\chi^g(x^{''},\vec{k}^{''}_\perp)&=& \frac{\sqrt{2\pi^{3/2}}}
{\sqrt{N_c \beta x^{''}(1-x^{''})}} e^{(\Delta^2-4\vec{k}^{'' 2})/8\beta^2}
\; .
\eeq
This model solves the problem of continuity, as shown in 
Fig.~\protect\ref{SDQCONT}

\newpage

\section{Rare $B\to K\ell^+\ell^-$ Decays}

   There is currently a great interest in the study of B decays for
study of the CKM mixing matrix as a test of electroweak models. The
exclusive $B\to K\ell^+\ell^-$ decay, a rare decay in that it is
forbidden for tree-level processes, is particularly interesting,
but the theoretical calculation requires hadronic form factors
involving nonperturbative physics. There have been many theoretical
studies in recent years\cite{JW,GK,Mel1,MN,HQ,Col,PB,BB,AKOS,Cas,Du}.
Since this decay of B-mesons involves a large energy transfer, it is
a natural for a light cone treatment, which has many advantages compared
to instant form relativistic quark model treatments\cite{JW,GK,Mel1,MN}.

   The effective Hamiltonian for the $b\to s\ell^+\ell^-$ decay process
in obtained after integrating out the heavy top quark and the $W^{\pm}$
bosons~\cite{GWS}: 
\beq\label{Hamil}
{\cal H}_{\rm eff}&=&\frac{4G_F}{\sqrt{2}}V_{tb}V^*_{ts}
\sum_i C_i(\mu)O_i(\mu),
\eeq
where $G_F$ is the Fermi constant, $V_{ij}$ are the
CKM matrix elements and $C_i(\mu)$ are the Wilson coefficients. 
The important operators $O_i(\mu)$ to the rare $b\to s\ell^+\ell^-$ 
decay are
\beq\label{OPE}
O_1&=& (\bar{s}_\alpha \gamma^\mu P_L b_\alpha)
(\bar{c}_\beta\gamma^\mu P_L c_\beta),\nonumber\\
O_2&=& (\bar{s}_\alpha \gamma^\mu P_L b_\beta)
(\bar{c}_\alpha\gamma^\mu P_L c_\beta),\nonumber\\
O_7&=&\frac{e}{16\pi^2}m_b(\bar{s}_\alpha\sigma_{\mu\nu}
P_R b_\alpha)F^{\mu\nu},\nonumber\\
O_9&=&\frac{e^2}{16\pi^2}(\bar{s}_\alpha\gamma^\mu P_L b_\alpha)
(\bar{\ell}\gamma_\mu\ell),\nonumber\\
O_{10}&=& \frac{e^2}{16\pi^2}(\bar{s}_\alpha\gamma^\mu P_L b_\alpha)
(\bar{\ell}\gamma_\mu\gamma_5\ell),
\eeq 
where $P_{L(R)}=(1\mp\gamma_5)/2$ is the chiral projection operator and
$F^{\mu\nu}$ is the electromagnetic interaction field strength tensor.
The Wilson coefficients $C_i(m_b)$ determined by the renormalization
group equations(RGE) from the perturbative value $C_i(M_W)$ are
given in the literature (see, for example~\cite{BM,Misiak}). After
integrating out the terms involving  $c\bar{c}$-loops and neglecting the
strange-quark mass, the resulting effective Hamiltonian corresponding
to Eq.~(\ref{Hamil}) is
\beq\label{Hll}
{\cal H}^{\ell^+\ell^-}_{\rm eff}&=&
\frac{4G_F}{\sqrt{2}}\frac{e^2}{16\pi^2}V^*_{ts}V_{tb}
\biggl[-\frac{2iC_7(m_b)m_b}{q^2}\bar{s}\sigma_{\mu\nu}q^\nu P_R b
\bar{\ell}\gamma^\mu\ell \nonumber\\
&+& C^{\rm eff}_9(m_b)\bar{s}\gamma_\mu P_L b\bar{\ell}\gamma^\mu\ell
+ C_{10}(m_b)\bar{s}\gamma_\mu P_L b\bar{\ell}\gamma^\mu\gamma_5\ell
\biggr],\nonumber\\
\eeq
where the effective Wilson coefficient $C^{\rm eff}_9$($\hat{s}$=$q^2/m^2_b$) 
is given in Refs~\cite{BM,Misiak,AMM,KMS,AKS,LW}, and has the form
\beq
\label{C9}
C^{\rm eff}_9(\hat{s}) &\equiv&
\tilde{C}^{\rm eff}_9(\hat{s}) + Y_{\rm LD}(\hat{s}),\nonumber\\
&=&C_9\biggl(1+\frac{\alpha_s(\mu)}{\pi}\omega(\hat{s})\biggr)
+Y_{\rm SD}(\hat{s})+Y_{\rm LD}(\hat{s}),
\eeq
where the functions $Y_{\rm SD}(\hat{s})$ is the one-loop
matrix element of $O_9$, $Y_{\rm LD}(\hat{s})$ describes
the long distance contributions due to the charmonium vector resonances 
via $B\to K(J/\psi,\psi',\cdots)\to K\ell^+\ell^-$, and $\omega(\hat{s})$
represents the one-gluon correction to the matrix element of $O_9$.

The parameters used
here, given by Refs.~\cite{AKS,LW}, are $m_t=175$ GeV, $m_b=4.8$
GeV, $m_c=1.4$ GeV, $\alpha_s(M_W)=0.12$, $\alpha_s(m_b)=0.22$,
$C_1=-0.26$, $C_2=1.11$, $C_3=0.01$, $C_4=-0.03$, $C_5=0.008$,
$C_6=-0.03$, $C_7=-0.32$, $C_9=4.26$, and
$C_{10}=-4.62$.

The long-distance contribution to $B\to K$ decay is contained in
the meson matrix elements of the bilinear quark currents appearing
in ${\cal H}_{\rm eff}$.
The matrix elements of the hadronic currents for $B\to K$ transition
can be parametrized in terms of hadronic form factors as follows:
\beq
\label{Jmu}
J^\mu &\equiv& \langle K|\bar{s}\gamma^{\mu}P_L b|B\rangle \nonumber \\
&=& \frac{1}{2}[F_{+}(q^{2})P^\mu + F_{-}(q^{2})q^\mu],
\eeq
and
\beq\label{JTmu}
J^\mu_T &\equiv& \langle K|\bar{s}i\sigma^{\mu\nu}q_\nu P_R b|B\rangle
\nonumber\\
&=& \frac{1}{2(M_B+M_K)}[q^2 P^\mu - (M^2_B-M^2_K)q^\mu] F_T(q^2),
\eeq
where $P=P_B+P_K$ and $q=P_B-P_K$ is the four-momentum
transfer to the lepton pair and $4m^{2}_{l}\leq q^{2}\leq(M_B-M_K)^2$.

  The transition amplitude for the $B\to K\ell^+\ell^-$ decay can be 
written as
\beq\label{TranA}
{\cal M} &=& \langle K\ell^+\ell^-|{\cal H}_{\rm eff}|B\rangle \nonumber\\
&=&\frac{4G_F}{\sqrt{2}}\frac{\alpha}{4\pi}V^*_{ts}V_{tb}
\biggl\{\biggl[C^{\rm eff}_9 J_\mu - \frac{2m_b}{q^2}C_7 J^T_\mu\biggr]
\bar{\ell}\gamma^\mu\ell \nonumber\\
&&\hspace{2.5cm}
+ C_{10}J_\mu \bar{\ell}\gamma^\mu\gamma_5\ell \biggr\},
\eeq
where $\alpha=e^2/4\pi$ is the fine structure constant.
The differential decay rate for $B\to K\ell^+\ell^-$ with nonzero 
lepton mass($m_\ell\neq 0$) is given by
\beq\label{DDR}
\frac{d\Gamma}{d\hat{s}}
&=&\frac{M^5_BG^2_F}{3\cdot2^9\pi^5}\alpha^2
|V^*_{ts}V_{tb}|^2\hat{\phi}^{1/2}
\biggl(1-4\frac{\hat{m}_\ell}{\hat{s}}\biggr)^{1/2}
\nonumber\\
&&\times
\biggl[\hat{\phi}\biggl(1+2\frac{\hat{m}_\ell}{\hat{s}}\biggr)
F_{T+} + 6\frac{\hat{m}_\ell}{\hat{s}}F_{0+} \biggr],
\eeq
where
\beq\label{DDR2}
F_{T+}&=&|C^{\rm eff}_9 F_+ - \frac{2C_7}{1+\sqrt{\hat{r}}}F_T|^2
+ |C_{10}|^2|F_+|^2,\nonumber\\
F_{0+}&=& |C_{10}|^2 [(1-\hat{r})^2|F_0|^2-\hat{\phi}|F_+|^2],
\nonumber\\
\hat{\phi}&=& (\hat{s}-1-\hat{r})^2-4\hat{r},
\eeq
with $\hat{s}=q^2/M^2_B$, $\hat{m}_\ell=m^2_\ell/M^2_B$,
$\hat{r}=M^2_K/M^2_B$, and
\beq
\label{Fo}
F_o(q^2) &=& F_+(q^2) + \frac{q^2}{M^2_B-M^2_K}F_-(q^2) \; 
\eeq
Note (Eqs.(\ref{DDR},\ref{DDR2})) that  only  $F_+(q^2)$ and $F_T(q^2)$ are 
necessary for the massless($m_\ell=0$) rare exclusive semileptonic 
$b\to s\ell^+\ell^-$, with $F_{o+}(q^2)$ not contributing.

   The differential branching ratio, $dBR(B\to K\ell^+\ell^-)/d\hat{s}$
 is obtained from
\beq
\label{DBR}
   \frac{dBR(B\to K\ell^+\ell^-)}{d\hat{s}} &=& 
\frac{d\Gamma(B\to K\ell^+\ell^-)}{\hat{s}} \frac{1}{\Gamma_{\rm tot}} \; '
\eeq
with $\Gamma_{\rm tot}=\frac{ M^5_BG^2_F}{64\pi^3}|V_{cb}|^2$
the total width.  
Another interesting observable, the longitudinal lepton
polarization asymmetry(LPA) is defined as
\beq
\label{LPA}
P_L(\hat{s})=\frac{d\Gamma_{h=-1}/d\hat{s}-d\Gamma_{h=1}/d\hat{s}}
{d\Gamma_{h=-1}/d\hat{s} +d\Gamma_{h=1}/d\hat{s}},
\eeq
where $h=+1(-1)$ denotes right (left) handed $\ell^-$ in the final state.
From Eq.~(\ref{DDR}), one obtains for $B\to K\ell^+\ell^-$
\beq
\label{LPA_BK}
P_L(\hat{s})=\frac{
2\biggl(1-4\frac{\hat{m}_\ell}{\hat{s}}\biggr)^{1/2}
\hat{\phi}C_{10}F_{+}
\biggl[F_+ {\rm Re}C^{\rm eff}_9 -
\frac{2C_7}{1+\sqrt{\hat{r}}}F_T\biggr] }
{ \biggl[\hat{\phi}\biggl(1+2\frac{\hat{m}_\ell}{\hat{s}}\biggr)
F_{T+} + 6\frac{\hat{m}_\ell}{\hat{s}}F_{0+} \biggr] } \; .
\eeq

The calculation of $B\to K\ell^+\ell^-$ is similar to the calculation
of ${\cal F}_{\pi}(\xi,x,t)$ described in the previous section.
An important difference is that in calculatin $F_+(q^2)$ and $F_T(q^2)$
the initial and final B-S amplitudes are different, so that in
contrast to Eqs.(\ref{jv}) and (\ref{jnv}) with $\chi$ in the initial
and final state, one needs a  $\chi_1$ and  $\chi_2$ for the initial b-meson
final s-meson. 

First, if one works in $q^+ = 0$ frame, with $q^2  = -\vec{q}^2_\perp < 0$
the weak form factors needed are of the form
\beq
\label{FP}
F_{+}(\vec{q}_{\perp}^{2}) &=&
\int^{1}_{0}dx\int d^{2}{\vec k}_{\perp}H_1(x,{\vec k}_{\perp},
{\vec k'}_{\perp})\phi_{2}(x,{\vec k'}_{\perp})\phi_{1}(x,{\vec k}_{\perp})
\eeq
and
\beq
\label{FT}
F_{T}(\vec{q}_{\perp}^{2}) &=&
\int^{1}_{0}dx\int d^{2}{\vec k}_{\perp} H_2(x,{\vec k}_{\perp},
{\vec k'}_{\perp})\phi_{2}(x,{\vec k'}_{\perp})\phi_{1}(x,{\vec k}_{\perp})
 \; ,
\eeq
with the known kinematic functions $H_1$ and $H_2$ given in Ref\cite{cjk01}.
The model light front Gaussian wave functions used are 
\beq
\label{Rad}
\phi(x,{\vec k}_{\perp}) &=&
\biggl(\frac{1}{\pi^{3/2}\beta^{3}}\biggr)^{1/2}
\exp(-\vec{k}^{2}/2\beta^{2}) \; ,
\eeq
normalized by $\int d^3k|\phi(x,{\vec k}_{\perp})|^2=1$
where $\vec{k}^2=\vec{k}^2_\perp + k^2_z$ and $k_z$ is defined as
\beq
\label{kz}
k_z &=& (x-\frac{1}{2})M_0 + \frac{m^2_q-m^2_{\bar q}}{2M_0}.
\eeq

   The calculations in Ref\cite{cjk01} are done in the purely longitudinal
frame with $q^+>0$ and ${\vec P}_{1\perp}={\vec P}_{2\perp}=0$, and thus
the momentum transfer square $q^2=q^+q^->0$ is timelike.
For the valence process, corresponding to Fig~\ref{highFock}(b), the 
calculation is carried out using the model l-f wave functions 
$\phi(x,{\vec k}_{\perp})$. For the nonvalence process, corresponding to 
Fig~\ref{highFock}(c), the final state B-S amplitude must be replaced
by the function
\beq
\label{GBK}
G_{BK}(x,{\vec k}_\perp) &\equiv&
\int \frac{dy}{\sqrt{y(1-y)}}\int d^2{\vec l}_{\perp} 
H_3((x,{\vec k}_{\perp};y,{\vec l}_{\perp})
{\cal K}(x,{\vec k}_{\perp};y,{\vec l}_{\perp}) \; ,
\eeq
where $H_3$ is a known function of the kinematic variables.
As discussed in the treatment of skewed quark distributions in the previous
section, in Ref~\cite{cjk01} $G_{BK}(x,{\vec k}_\perp)$ is taken as a
constant, and the reliablity of this approximation is checked by examining 
the frame-independence of the numerical results.
 
\begin{figure}[p]
\centerline{\psfig{figure=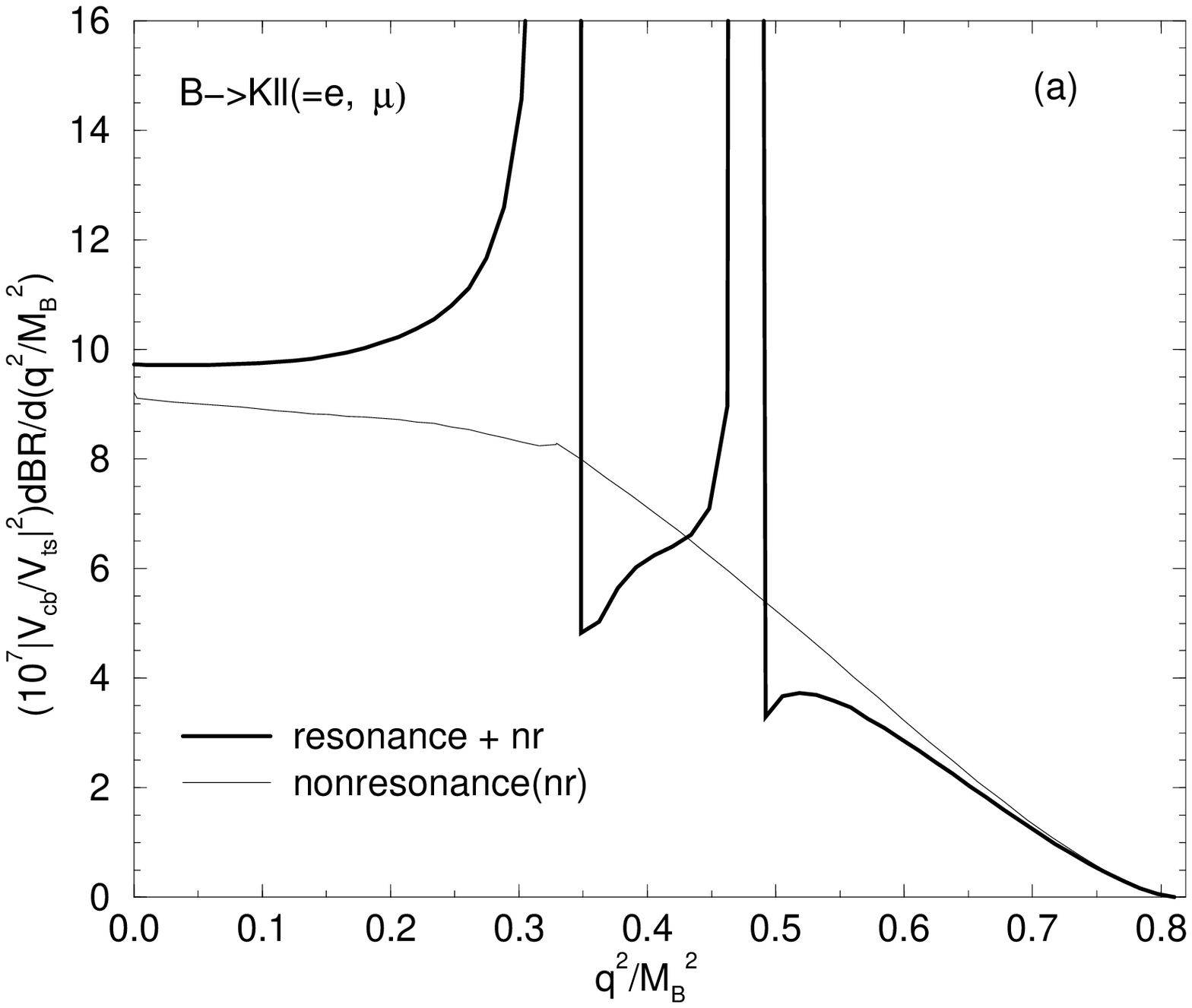,height=6.5cm,width=9cm}}
\centerline{\psfig{figure=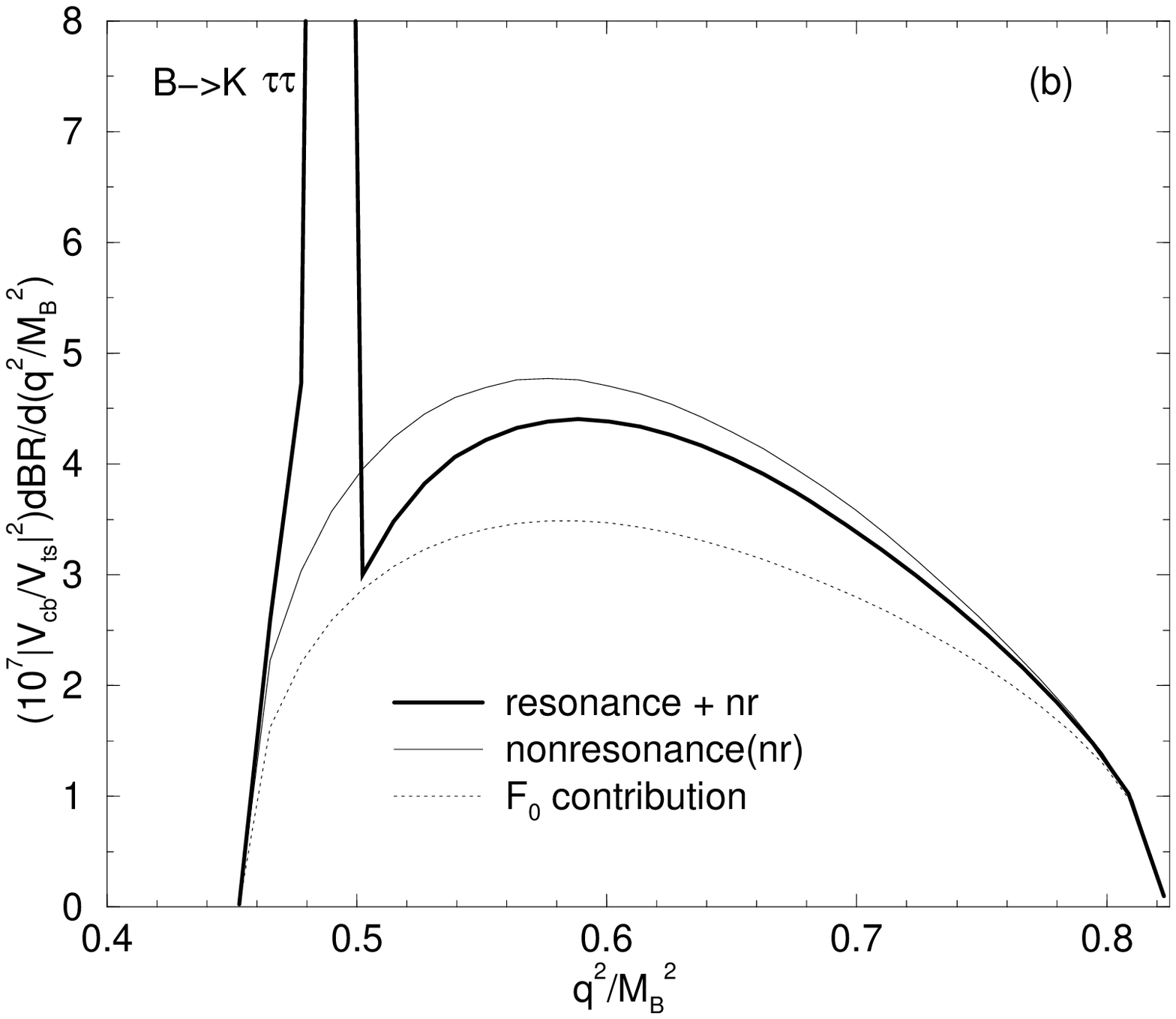,height=6.5cm,width=9cm}}
\caption{ The branching ratios for
$B\to K\ell^+\ell^-$(a) and $B\to K\tau^+\tau^-$(b)
transitions. The thick(thin) solid line represents the result
with(without) LD contribution($Y_{LD}$) to
$C^{\rm eff}_9$ in Eq.~(\protect\ref{DDR2}). The dotted line
in (b) represents the $F_0(q^2)$ contribution to the total branching
ratio of $\tau$ decay.\label{fig_BKrate}}
\end{figure}

\begin{figure}[p]
\centerline{\psfig{figure=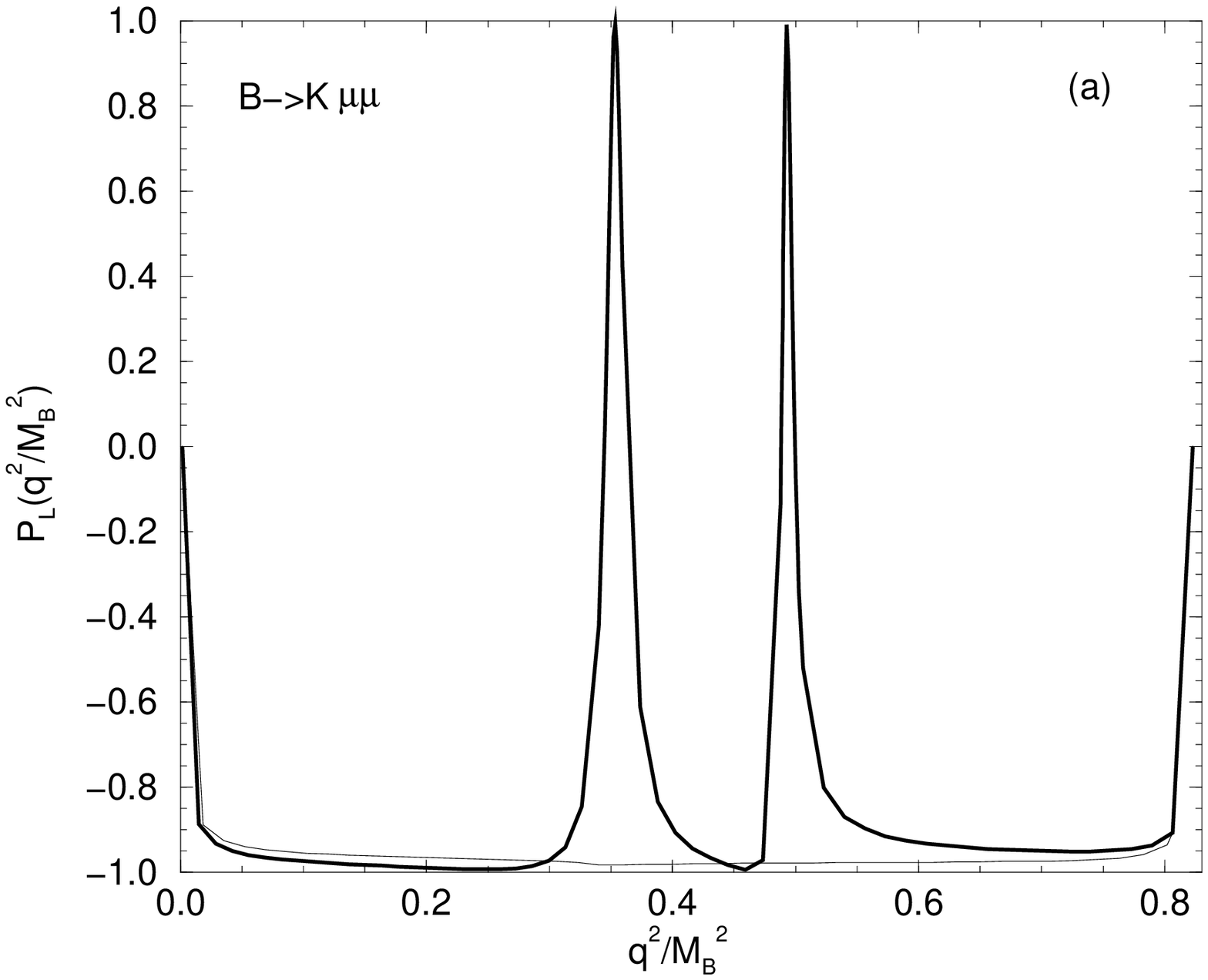,height=6.5cm,width=9cm}}
\centerline{\psfig{figure=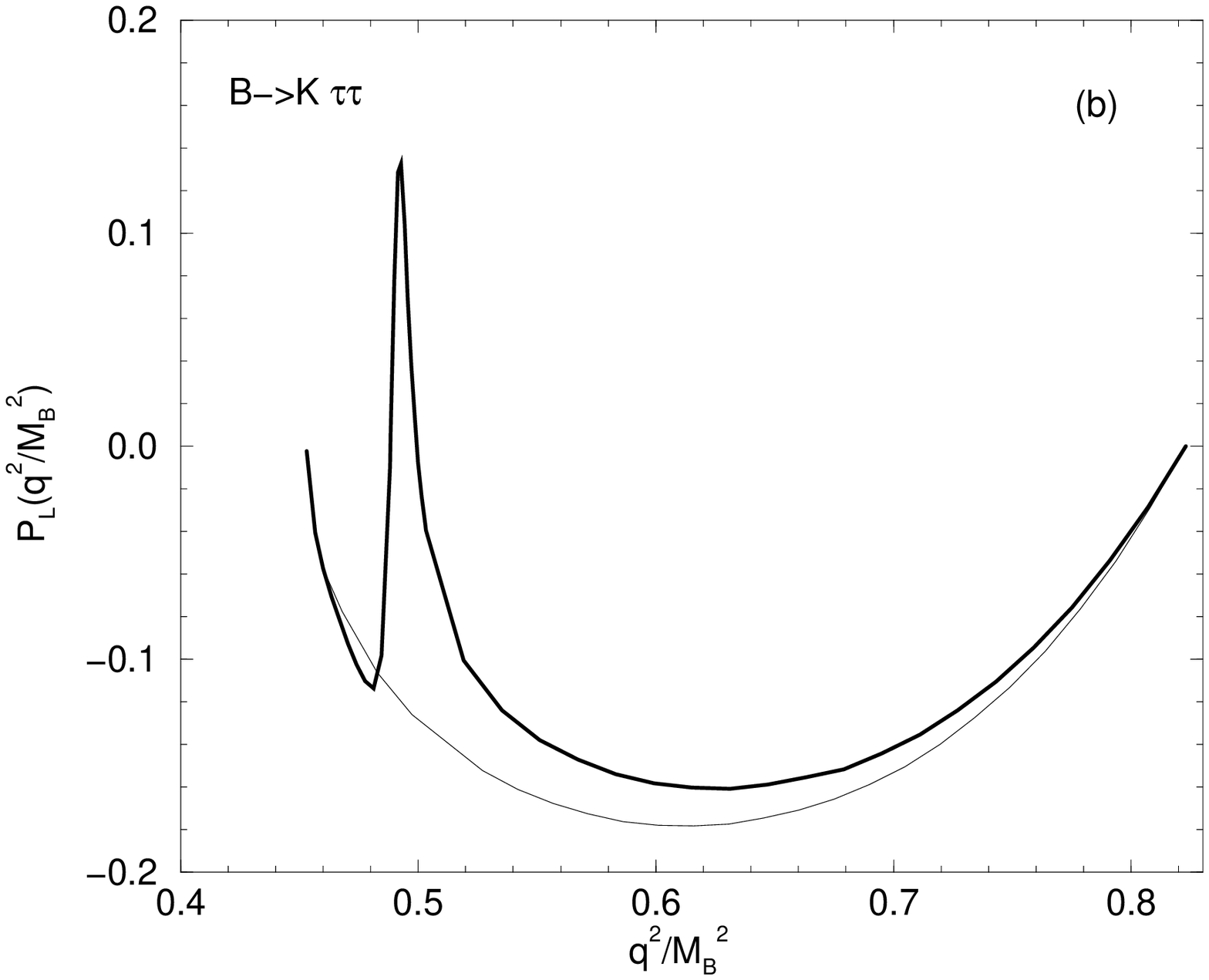,height=6.5cm,width=9cm}}
\caption{ The longitudinal lepton polarization asymmetry $P_L(\hat{s})$
for $B\to K\ell^+\ell^-$(a) and $B\to K\tau^+\tau^-$(b)
transitions. The same line code is used as in Fig.~\protect\ref{fig_BKrate}.
\label{fig_LPA}}
\end{figure}

 The results with this model for the differential branching ratios,
 with $m_\ell=0$ and $m_{\tau}$=1.777 GeV, 
for $B\to K\ell^+\ell^-(\ell=e,\mu)$ 
are shown in Fig.~\ref{fig_BKrate}(a)
and for $B\to K\tau^+\tau^-$ in Fig.~\ref{fig_BKrate}(b), respectively.
The thick(thin) solid line represents the result with(without)
the LD contribution($Y_{LD}(\hat{s})$) to $C^{\rm eff}_9$ given by
Eq.~(\ref{C9}).
One can see that the pole contributions clearly overwhelm the branching
ratio near $J/\psi(1S)$ and $\psi'(2S)$ peaks, however, suitable
$\ell^+\ell^-$ invariant mass cuts can separate the LD contribution
from SD one away from these peaks.
This divides the spectrum into two distinct regions~\cite{Hew,AGM}:
(i) low-dilepton mass, $4m^2_\ell\leq q^2\leq M^2_{J/\psi}-\delta$,
and (ii) high-dilepton mass,
$M^2_{\psi'}+\delta\leq q^2\leq q^2_{\rm max}$, where $\delta$ is to
be matched to an experimental cut.
The numerical results for the non-resonant branching ratios(assuming
$|V_{tb}|\simeq 1$) are $4.96\times 10^{-7}|V_{ts}/V_{cb}|^2$
for $B\to K\ell^+\ell^-$ ($\ell=e,\mu$) and
$1.27\times 10^{-7}|V_{ts}/V_{cb}|^2$ for
$B\to K\tau^+\tau^-$, respectively.
While the data by CLEO Collaboration~\cite{Bfactory}
reported the branching ratio ${\rm Br}(B\to K e^+e^-)<1.7\times 10^{-6}$,
the data by Belle Collaboration(K. Abe et al.)~\cite{Bfactory}
reported ${\rm Br}(B\to K e^+e^-)<1.2\times 10^{-6}$ and
${\rm Br}(B\to K\mu^+\mu^-)=(0.99^{+0.39+0.13}_{-0.32-0.15})\times 10^{-6}$,
respectively. The comparison of the results of this model with other
models is given in detail in Ref~\cite {cjk01}

The longitudinal lepton polarization asymmetry(LPA) in this model for 
$B\to K\mu^+\mu^-$ and $B\to K\tau^+\tau^-$ as a function of $\hat{s}$, 
respectively, in Figs.~\ref{fig_LPA}(a) and~(b). 
The results with the LD contributions are shown be the thick solid line,
and without by the thin solid line. Further discussion of these results and
comparison with other calculations is found in Ref~\cite{cjk01}.

In conclusion, the rare exclusive semileptonic $B\to K\ell^+\ell^-$ 
($\ell=e,\mu$ and $\tau$) decays are found to be useful for studies of
nonperturbative form factors, and the light cone approach is most valuable
for the theoretical models.

\newpage

\section{Meson Light Cone Wave Functions}

   The structure of mesons is of great interest for the study of QCD.
With a simple two-body configuration ${\bar q}q$ being the lowest Fock 
state, for certain reactions the pion can be a relatively simple system 
for theoretical investigations.  As discussed in Sec. 4, for high momentum 
transfer processes the PQCD treatment is adequate. At low and medium 
momentum transfers nonperturbative effects dominate. In the present
section the QCD sum rule method for determining the wave function
of the $\pi$ and $\rho$ mesons is reviewed.
The pion wave function is of particular interest as there are now
experimental studies of the light cone wave function, which is
discussed in this section.

  In Sec.4 the  Bethe-Salpeter amplitude for a two-body bound state,
$\Psi(z_1,z_2) = \\
 <0|T[\Psi(z_1)\Psi(z_2)]|M>$ in instant form coordinate
space, was discussed, with the light cone B-S amplitude, 
$\Psi(x_1,x_2,\vec{k}_\perp)$, given as the
solution to the l-c B-S equation, Eq.(~\ref{lcBS}). In order to obtain
the correct physical B-S amplitude, also called the light cone wave function,
one needs a kernel $( K(x_1,x_2,\vec{k}_\perp,y_1,y_2,\vec{l}_\perp)$ which 
gives a satisfactory representation of the field theory being considered. 
In the applications discussed in previous sections models were used. In the 
next subsection we discuss the use of QCD sum rules to estimate the pion 
and rho meson wave functions. Theoretical and experimental work on the direct
measurement of the ${\bar q}q$ component of the pion wave function is
discussed in the following subsection.

\subsection{QCD Sum rules for the Pion and Rho Meson Wave Function}

  We first discuss the pion wave function is some detail, and then give the 
results for the rho.  The light cone wave function of a $\pi^+$ meson can be 
obtained from the gauge-invariant matrix element
\beq
\label{lfwfo}
  \Psi_\mu(z\cdot p,z^2) &=& <0|{\bar d}(0)\gamma_\mu{\cal P}
e^{\int_o^1 dy^\mu A_\mu} \gamma_5 u(z)|\pi^+(p)> \;,
\eeq
where (u,d) are the up and down quark fields and $A_\mu = A_\mu^n\lambda^n/2$
is the QCD gluonic color field. We shall work in the fixed point gauge, 
with the gauge condition $x_\mu A^\mu = 0$, so that the operator
$e^{\int_o^1 dy^\mu A_\mu} = 1$. The method of QCD sum rules\cite{svz} 
utilizes an operator product expansion, which leads in a natural way to a 
twist expansion of the wave function. We briefly review QCD sum rule procedure

\subsubsection{Brief Review of the QCD Sum Rule Method}

 The method of QCD sum rules makes use of a correlator defined in terms 
of a composite field operators $\eta_a, \eta_b$
\beq
\label{qcd1}
\Pi_{ab} (p) &=& i\int d^4 x \; e^{iq \cdot x} 
 <0 \mid T[\eta_a(x)\eta_b(0)] \mid 0> \;.
\eeq
For mesons the $\eta(x)$ can be taken in the form
\beq
\label{qcd2}
            \eta_a (x) &=& \bar{q}(x) \Gamma_a q(x) \; ,
 \eeq
with $\Gamma_a$ chosen so that the operator $\eta_a$ creates states with
the quantum numbers of the meson under consideration. E.g., for a scalar,
pseudoscalar or vector meson $\Gamma_a =1, \; \gamma_\mu\gamma_5$ 
or $\gamma_\mu$, respectively, are possible choices.

 If the correlator satisfies an unsubtracted 
dispersion relation (in Euclidean space), then $\Pi_{ab}$ can be expressed as
\beq
\label{qcd3}
  \Pi_{ab}(p) &=& \frac{1}{\pi}\int_0^\infty ds \frac{Im\Pi_{ab}(s)}{s+p^2} 
\nonumber \\
              &=& \frac{N_{ab}}{M^2 +p^2} + 
  \frac{1}{\pi} \int_{s_o}^\infty ds \frac{Im\Pi_{ab}(s)}{s+p^2} \; ,
\eeq
where M is the mass of the stable meson, $N_{ab}$ is a constant, and 
$s_o$ is the parameter
for the beginning of the continuum contribution to the dispersion relation.
The QCD sum rule is obtaind  by equating the Borel transformation, 
${\cal B}_{-p^2\rightarrow M_B^2}$, of the phenomenological correlator,
Eq.(\ref{qcd3}), to the QCD evaluation of $\Pi_{ab}$,
\beq
\label{qcd4}
  \Pi_{ab}(M_B) &=& N_{ab}e^{-M^2/M_B^2} +
 \frac{1}{\pi} \int_{s_o}^\infty ds  e^{-s/M_B^2} Im\Pi_{ab}(s) \nonumber \\
                &=& \Pi_{ab}^{QCD}(M_B) \; ,
\eeq
with $M_B$ called the Borel mass.
The Borel transformation reduces the importance of the continuum contribution,
and results in a rapidly diminishing size of the QCD diagrams with
increasing dimension. The QCD side of Eq.{\ref{qcd4} is of the form 
\beq
\label{qcd5}
 \Pi_{ab}^{QCD}(M_B) &=& \Pi_{ab}^{PQCD}(M_B) +\frac{\Sigma_n c_n <0_n>}
{M_B^n} \;,
\eeq
where the $c_n$ are constants obtained by evaluating the QCD diagrams and
the $<0_n>$ are vacuum condensates of dimension $n$. In order of increasing
dimension the most important three condensates are
\beq
\label{qcd6}
        <O_3> &=& <\bar{q}q> \; {\rm quark \;\;condensate} \nonumber \\
      <0_4> &=& <(G^{\mu\nu}_a)^2> \; {\rm gluon \; \; condensate} \nonumber \\
      <0_5> &=& <\bar{q}\sigma^{\mu\nu}G_{\mu\nu}q> \; {\rm mixed \; \;
 condensate} \; .
\eeq
The quark condensate is rather accurately known, with $-(2\pi)^2<\bar{q}q>
\simeq$ 0.55 GeV$^3$. A typical value for the gluon condensate is
$<g_s^2 G^2> \simeq$ .47 GeV$^4$, and for the mixed condensate
$< g_s \bar{q}\sigma^{\mu\nu}G_{\mu\nu}q> \equiv  -m_o^2 <\bar{q}q>$, with
$m_o^2 \simeq$ 0.8 GeV$^2$. 

\subsubsection{QCD Sum Rules For the $\pi$ Light Cone Wave Function}

  A QCD sum rule analysis of $\Psi_\mu(x\cdot p,x^2)$ to extract an
expansion of the pion l-c wave function with nonperturbative QCD effects
was carried out in Ref.~\cite{cz82}. See Refs.~\cite{cz84,st97} for 
reviews. A study of the l-c wave function vs. light front quark models
has also been done using QCD sum rule techniques\cite{bj97}

   Using the fixed point gauge and expanding u(z) about the origin,
one obtains for \\$\Psi_\mu(z\cdot p,z^2)$ (Eq.(\ref{lfwfo})) 
\beq
\label{cz1}
 <0|{\bar d}(0)\gamma_\mu \gamma_5 u(z)|\pi^+(p)> &=& 
  \Sigma_n\frac{(-1)^n}{n!} <0|d(0)\gamma_\mu \gamma_5 (z_\nu D^\nu)^n
|\pi(p)> \nonumber \\
          &=& p_\nu \hat{\phi}_\pi(z\cdot p,z^2),
\eeq.
with $D^\nu = \partial^\nu -igA^\nu$ and $\hat{\phi}$ proportional to the
l-c wave function.

This expression is evaluated using a version of QCD sum rules. Using
$\eta_{\pi^+ \mu} (x)=\bar{d}(x)\gamma_\mu\gamma_5u(x)$ for the pion
composite field operator, the n$^{th}$ term in Eq.(\ref{cz1}) is given by 
the vacuum matrix elements
\beq
\label{cz2}
       I_n(p^2) &\equiv& \frac{i}{(-z\cdot p)^{n+2}}\int dx e^{ip \cdot x}
 <0|\bar{d}(x)\not\!{z}\gamma_5(z\cdot D)^n u(x) \bar{u}(0)\not\!{z}
 \gamma_5 d(0)|0> \; .
\eeq
By evaluating the diagrams up to dimension 6, illustrated in 
Fig.~\ref{qcdsum1}, Chernyak and Zhitnitsky derived the expansion
\begin{figure}
\begin{center}
\epsfig{file=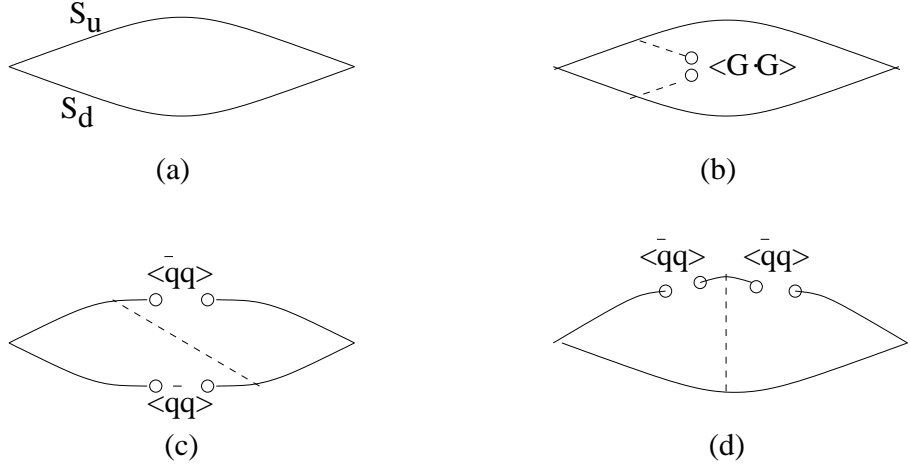,width=12cm}
\caption{Diagrams for QCD calculation of $\phi_\pi$ up to D=6. (a) is the
PQCD process, (b) is the D=4 gluon condensate process, (c), (d) are the
processes with two quark condensates}
{\label{qcdsum1}}
\end{center}
\end{figure}
for the Borel transform of $I_n(p^2)$,
\beq
\label{cz3}
  \frac{4\pi}{M_B^2} I_n(M_B^2) &=& \frac{3}{(n^2 +4n + 3)} +
   \pi\alpha_s \left(\frac{<0|G^2|0>}{3M_B^4} + \frac{4}{81 \pi^2}
<0|(2\pi)^2 \bar{q}q|0>^2 \right) +... \;.
\eeq

The light cone wave function is obtained from $ \hat{\phi}_\pi(z\cdot p,z^2)$
of Eq.(\ref{cz1}) by letting $\gamma^\mu \rightarrow \gamma^+ \equiv
 (\gamma^0 + \gamma^3)/\sqrt{2}$ and $z^+=z_\perp =0$. The light cone
wave function is then obtianed from $\Psi(z^-p^+,0)$ in the gauge with
$A^+ = 0$ :
\beq
\label{cz4}
     \phi_\pi(x) &=& \int_\infty^\infty \frac{dz^-}{2\pi}e^{iz^- x p^+}
  \Psi(z^-p^+,0) \; .
\eeq

The CZ wave function\cite{cz82}, with the normalization 
$\int dx \phi(x) = f_\pi$, with the pion decay constant  $f_\pi=$
133 MeV, was obtained using the values of the condensates given above.
Within the errors in the method an approximate form for the CZ result
often used is
\beq
\label{cz5}
    \phi_\pi^{CZ}(\xi)|_{Q=.5 {\rm GeV}} &=& 5 x(1-x)(1-2x)^2 \; , 
\eeq
with units of $f_\pi/\sqrt(3/2)$
This can be compared to the asymptotic light cone wave function\cite{lb80},
also with units of $f_\pi/\sqrt(3/2)$
\beq
\label{cz6}
 \phi_\pi^{AS}(x)|_{Q^2 \rightarrow \infty} = x(1-x) \; .
\eeq

\subsubsection{QCD Sum Rules For The Light Cone Wave Function of the $\rho$}

  Treatment of the rho meson, with the composite field operator $\eta^\rho_\mu$
is quite different for the longitudinal ($\eta^\rho_{\mu=0}$) and the
transverse ($\eta^\rho_{\mu=\pm 1})$ currents, for $\rho_L$ and $\rho_\perp$,
respectively.

 For $\rho_L$ the QCD sum rule treatment of Ref~\cite{cz82} is the same as
for the $\pi$, with the quantities $I_n(p^2)$ given by the expression in
Eq.(\ref{cz2}) with the operators $\gamma_5$ removed. The resulting expansion
of the l-c wave function, using the analog of the diagrams in 
Fig.~\ref{qcdsum1}, in units of 3f$_\rho$ is approximately
\beq
\label{cz7}
   \phi_{\rho ,L}^{CZ}(\xi)|_{Q=.5 {\rm GeV}} &=& x(1-x)(0.7 +1.5(1-2x)^2)
 \; , 
\eeq
For $\rho_\perp$  the quantities corresponding to the $I_n(p^2)$ given by 
the expression in Eq.(\ref{cz2}) for the $\pi$ are obtained by replacing
$z_\nu\gamma_\nu\gamma_5$ by $\sigma_{\mu\nu}z_\nu$ in Eq.(\ref{cz2})
(with a sum over $\mu$, of course). The result of the analysis\cite{cz82}
for the transverse rho l-c wave function, in units of 3f$_\rho$ is 
approximately
\beq
\label{cz7a}
   \phi_{\rho ,\perp}^{CZ}(\xi)|_{Q=.5 {\rm GeV}} &=& 5(x(1-x))^2
 \; , 
\eeq

In Fig.~\ref{qcdsum2} the CZ pion l-c pion wave function is compared 
to the asymptotic l-c pion wavefunction, in units of $f_\pi/\sqrt{3/2}$, 
for the pion and the transverse rho l-c CZ wavefunction 
\begin{figure}
\begin{center}
\epsfig{file=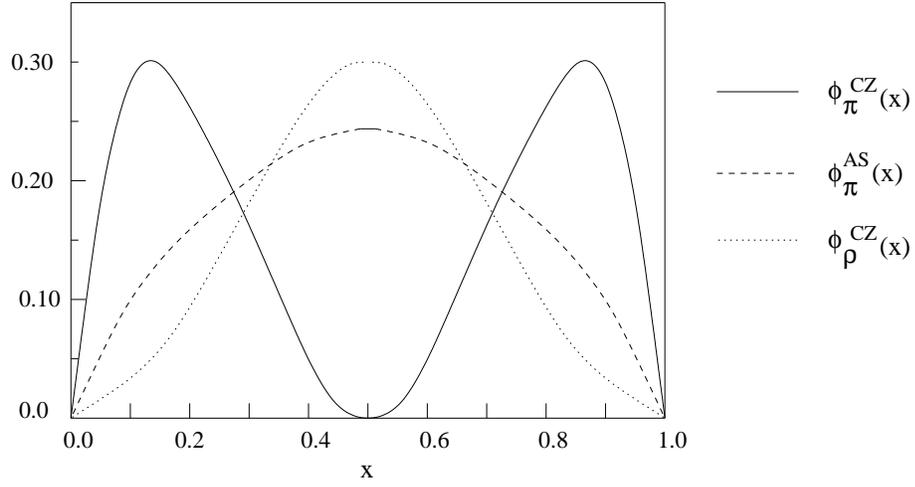,width=12cm}
\caption{Comparison of the CZ with the asymptotic $\pi$ light cone wave 
function, in units of $f_\pi/\sqrt{2}$, and the transverse $\rho$ CZ
l-c wave function.}
{\label{qcdsum2}}
\end{center}
\end{figure}

\section{Factorization and Experimental Study of Light Cone Wave Functions}

   An essential aspect of testing hadronic and nuclear structure by high
energy reactions is that amplitudes for such reactions factorize. Consider
Fig.~\ref{qcdsum3} with a schematic high energy reaction on a target with
\begin{figure}
\begin{center}
\epsfig{file=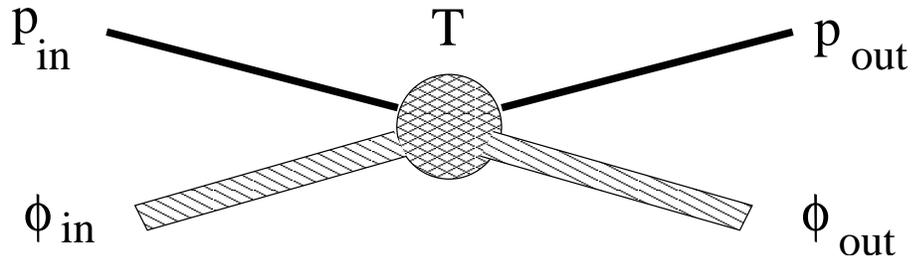,width=12cm}
\caption{High-energy reaction of a projectile on a target (in) leading to
a final state (out).}
{\label{qcdsum3}}
\end{center}
\end{figure}
wave function $\phi_{in}$ leading to a final state with wave function
 $\phi_{out}$. The factorization theorem states that the reaction
amplitude satisfies the schematic relationship
\beq
\label{cz8}     
  {\cal M}(in,p\rightarrow out,p') &=& \int_0^1dx\int_0^1ddy
\Phi_{in}(x,p_i)T(x,y,Q^2)\phi_{out}(y,p_f) \; ,
\eeq
with $\phi_{in},\phi_{out}$ the initial and final target wave functions,
and $T$ represents the parton hard scattering.
Factorization is important for the extraction of information for many 
reactions. For a discussion with references see, e.g., Ref.~\cite{st97}.
The coherent production of dijets with a high energy pion projectile,
which we discuss next, is of particular interest because of a recent
experimental measurement.

\subsubsection{Coherent Diffractive Pion Production of Dijets}

  The concept of diffractive dissociation is very old in nuclear/particle
physics\cite{fp56}, and was extended to QCD in Ref.~\cite{bbgg81}. The
essential idea for pion production of dijets is that for a pion projectile 
on a nuclear target at high energy, the short-range color zero  $\bar{q}q$ 
configuration will traverse the nucleus with little interaction, while the
other components will be absorbed. Therefore the $\bar{q}q$  component is
filtered out and leaves the nucleus to form two jets.
\begin{figure}
\begin{center}
\epsfig{file=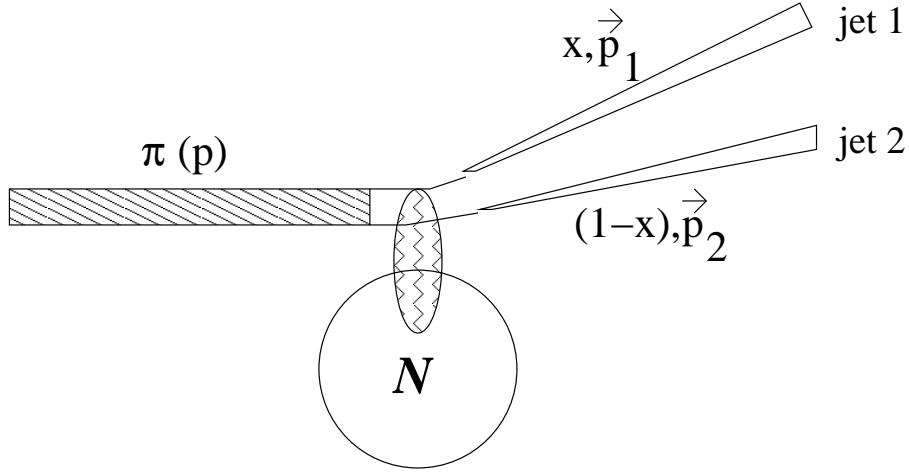,width=12cm}
\caption{Coherent diffractive dissociation of a pion projectile into two
jets with a nuclear (${\cal N}$) target}
{\label{qcdsum4}}
\end{center}
\end{figure}

   The $(\pi,{\rm dijet}$) amplitude\cite{fms93}, for high energies with
the forward  $\bar{q}q$ scattering amplitude $f(b^2) \simeq s\sigma(b^2)$,
where $\sigma(b^2)$ is the total cross section as a function of impact 
parameter $b$, with this coherent diffractive dissociation production 
scenerio is
\beq
\label{cz9}
  {\cal M} &=& \int db^2 \phi_\pi(x,\vec{b})\frac{s}{2}\sigma(b)
e^{i\vec{k}_\perp\cdot \vec{b}} \; ,
\eeq
where $\vec{b}$ is the impact parameter, $\sigma(b)$ is the total
forward cross section of the valence diquark $\bar{q}q$ system,
and the scattering of the final jets has been neglected, so that
their wave function is a plane wave with the relative transverse
momentum variable $\vec{k}_\perp$. This is a special case of the
factoriation theorem, Eq.(\ref{cz8}). Recently, there has been a theoretical
study of photoproduction of high energy dijets with nuclear 
targets\cite{braun02} which within the approximations leading to 
Eq.(\ref{cz9}) is the reaction
\beq
\label{cz10}
   \gamma + {\cal N} & \rightarrow & (\bar{q}q)+ {\cal N} \; ,
\eeq
so that the pion valence wave function can also be studied with this
reaction.

\subsubsection{Experimental Study of the Pion Light Cone $\bar{q}q$ Component}

   A direct measurement of the light cone meson wave function can be
carried out through a high-energy coherent diffractive dissociation
of the meson. This was recently done at Fermilab experiment E791\cite{E791}.
The experiment used a 500 GeV $\pi^-$ beam with a platinum target.
\begin{figure}[ht]
\vspace*{10 cm}
\includegraphics{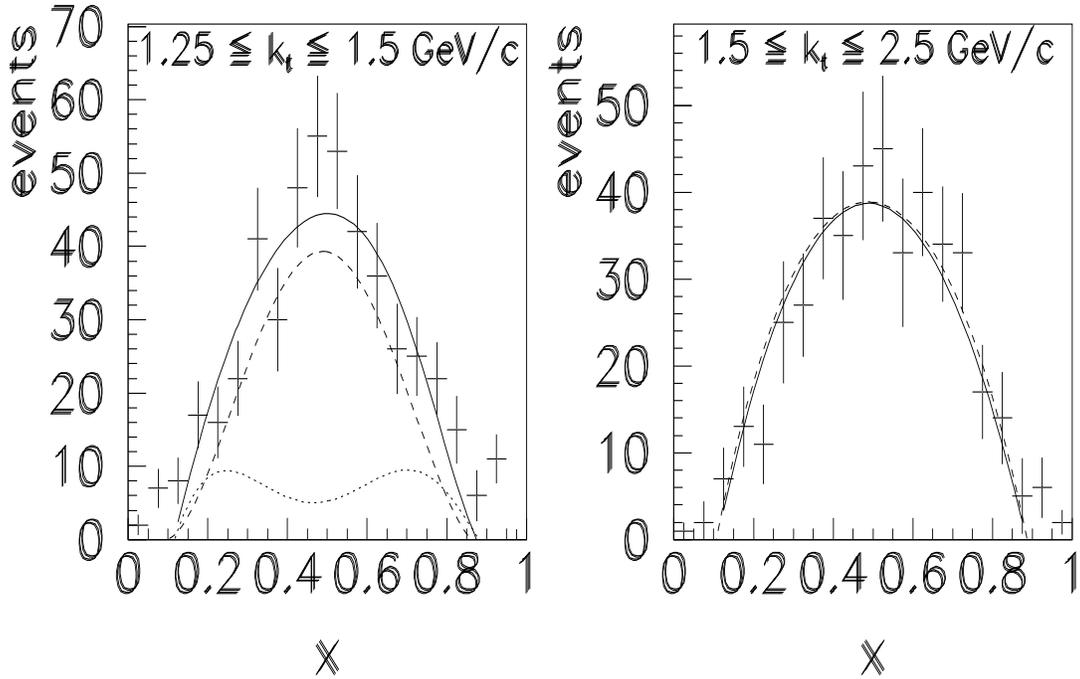}
\caption{x distribution of dijets in theE791 ${\cal N}(\pi,{\rm dijet})
{\cal N}$ reaction. The dashed line is the asymptotic, the dotted line 
the CZ and thesolid line is a combination of the CZ and asymptotic l-c
wave functions.}
{\label{qcdsum5}}
\end{figure} 
The diffractive dijets were identified as having their yield depend on
$q_\perp$, the transverse momentum transfered to the nucleus, by
\beq
\label{cz11}
            N_{jet} & \sim & e^{-<b^2> q_\perp^2} \; ,
\eeq
where $<b^2> = R_{\cal N}^2/3$ is the effective impact parameter squared.
The x value of the jet (see Fig.~\ref{qcdsum4}) is obtained from
\beq
\label{cz12}
          x_{exp} &=& \frac{p_{jet1}}{p_{jet1} + p_{jet2}} \; .
\eeq

The results of the experiment are shown in Fig.~\ref{qcdsum5}.
The fit to the data, the solid line in Fig~\ref{qcdsum5} is made with a
squared wave function that is a linear combination of the CZ and AS wave
functions
\beq
\label{cz13}
   \phi^2(x) &=& a_{AS}\phi_\pi^{AS}(x)^2 + a_{CZ} \phi_\pi^{CZ}(x)^2.
\eeq
Using the relationship for the virtuality of the dijets 
\beq
          Q^2 & \simeq & \frac{k_\perp^2}{x(1-x)} \; ,
\eeq
for $k_\perp >$ 1.5 GeV $Q^2 > \sim$ 10 GeV$^2$, and the $\phi_{AS}$
gives a good fit to the data, while in the lower $1.25 \leq 1.5$ bin
the CZ component makes a sizable contribution.

\section{Conclusions}

   The methods of light cone field theory and light front quantum mechanics
play a vital role in the extraction of information on the structure of 
hadrons and nuclei. It has been applied in recent years to hadronic form
factors and a variety of reactions, with a great deal of work on the
transition from nonperturbative to perturbative QCD treatments of hadronic
form factors and deeply virtual reactions with photon projectiles.
The use of light cone wave functions is important for extracting
information from the decay of heavy-quark hadrons.
Quite recently coherent diffractive dissociation and related reactions
have begun to provide direct measurements of light cone wave functions.
This is an exciting field for theory and experiment, and will continue
to provide important informationabout the nature of Quantum Chromodynamics.
\vspace{1cm}

{\bf ACKNOWLEDGEMENTS} The author would like to acknowledge many helpful
discussions with Matthias Burkhardt, Mikkel Johnson, and Stephan Pate,
coorganizers of the Light cone 2002 International Workshop, as well as
many of those attending the workshop. This work was supported in part by
the NSF grant PHY-00070888 and in part by the DOE contract W-7405-ENG-36.


\end{document}